%% file: main.tex
\documentclass[pmlr]{jmlr}% new name PMLR (Proceedings of Machine Learning)

\usepackage{longtable}% for long tables

 \usepackage{caption} 
 \usepackage{booktabs}
 \usepackage{multirow}
 \usepackage{colortbl}

 \usepackage{fancyhdr}
 
\definecolor{lightorange}{HTML}{FFCC99}
\usepackage[load-configurations=version-1]{siunitx} % newer version
\makeatletter
\def\set@curr@file#1{\def\@curr@file{#1}} %temp workaround for 2019 latex release
\makeatother

\theorembodyfont{\upshape}
\theoremheaderfont{\scshape}
\theorempostheader{:}
\theoremsep{\newline}

\jmlrvolume{182}
\jmlryear{2023}
\jmlrworkshop{Machine Learning for Healthcare}
\definecolor{forestgreen}{rgb}{0.13, 0.55, 0.13}

\newcommand\granularvariation{granular racial variation}
\newcommand\coarsegroup{coarse race group}
\newcommand\granulargroup{granular race group}
\newcommand\riskscore{\hat y}
\newcommand\outcome{y}
\newcommand\features{X}
\newcommand\Agranular{A^{(g)}}
\newcommand\Acoarse{A^{(c)}}

\usepackage[symbol]{footmisc}
\renewcommand{\thefootnote}{\fnsymbol{footnote}}

\title[Coarse race data conceals disparities in clinical risk score performance]{Coarse race data conceals disparities in clinical risk score performance}

\author{\Name{Rajiv Movva*}
        \Email{rmovva@cs.cornell.edu} \\ 
        \addr Cornell Tech 
        \AND
        \Name{Divya Shanmugam*}
        \Email{divyas@mit.edu} \\ 
        \addr Massachusetts Institute of Technology
        \AND
        \Name{Kaihua Hou}
        \Email{houk@berkeley.edu} \\ 
        \addr UC Berkeley 
        \AND
        \Name{Priya Pathak}
        \Email{pp2841@cumc.columbia.edu} \\ 
        \addr Columbia University Medical Center
        \AND
        \Name{John Guttag}
        \Email{guttag@csail.mit.edu} \\ 
        \addr Massachusetts Institute of Technology
        \AND
        \Name{Nikhil Garg}
        \Email{ngarg@cornell.edu} \\ 
        \addr Cornell Tech 
        \AND
        \Name{Emma Pierson}
        \Email{emma.pierson@cornell.edu} \\ 
        \addr Cornell Tech 
} 

\begin{document}

\maketitle
% comment out below line to bring back header + footer.
\thispagestyle{plain}

\def\thefootnote{*}
\footnotetext{Indicates co-first authorship.}
\def\thefootnote{\arabic{footnote}}

\begin{abstract}
  Healthcare data in the United States often records only a patient's coarse race group: for example, both Indian and Chinese patients are typically coded as ``Asian.'' 
  It is unknown, however, whether this coarse coding conceals meaningful disparities in the performance of clinical risk scores across granular race groups. 
  Here we show that it does.
  Using data from 418K emergency department visits, we assess clinical risk score performance disparities across 26 granular groups for three outcomes, five risk scores, and four performance metrics. 
  Across outcomes and metrics, we show that the risk scores exhibit significant granular performance disparities \textit{within} coarse race groups. 
  In fact, variation in performance within coarse groups often \textit{exceeds} the variation between coarse groups.
  We explore why these disparities arise, finding that outcome rates, feature distributions, and relationships between features and outcomes all vary significantly across granular groups. 
  Our results suggest that healthcare providers, hospital systems, and machine learning researchers should strive to collect, release, and use granular race data in place of coarse race data, and that existing analyses may significantly underestimate racial disparities in performance. 
\end{abstract}

\input{mlhc-camera-ready-template/01_introduction}

\input{mlhc-camera-ready-template/02_related_work}
\input{mlhc-camera-ready-template/03_methods}
\input{mlhc-camera-ready-template/04_describing_disparities}

\input{mlhc-camera-ready-template/05_explaining_disparities}
\input{mlhc-camera-ready-template/06_discussion}

\paragraph{Acknowledgements}  We thank Leo Celi, Serina Chang, Alistair Johnson, Aniruddh Raghu, and Kenny Peng for helpful comments. This research was supported by a Google Research Scholar award, NSF CAREER \#2142419, a CIFAR Azrieli Global scholarship, a LinkedIn Research Award, Wistron Corporation, a Future Fund Regrant, a Meta Research Award, the Abby Joseph Cohen Faculty Fund, a Cornell Tech Urban Tech Hub grant, and NSF GRFP DGE \#2139899.

\pagebreak

\bibliography{mlhc-camera-ready-template/zotero_cr}

\appendix

\input{mlhc-camera-ready-template/99_appendix}

\end{document}

%% file: mlhc-camera-ready-template/01_introduction.tex
\section{Introduction}

Despite large and persistent racial health disparities, race data in United States health records are often incorrect, incomplete, or missing altogether \citep{hahn_state_1992, klinger_accuracy_2015,polubriaginof_challenges_2019, jarrin_validity_2020}. Even when race is recorded, it often reflects a patient's \emph{coarse} race group, which combines several \emph{granular} groups into a single category \citep{hanna_towards_2020,borrell_race_2021, lett_conceptualizing_2022}. Past work has shown that coarse race categories can obscure consequential medical differences: for instance, diabetes is nearly twice as common in Indian patients compared to Chinese patients \citep{vicks_prevalence_2022}, which has motivated a wealth of research specific to the treatment and diagnosis of diabetes in each group \citep{sabanayagam_diagnosis_2015,ali_preventing_2020,mao_efficacy_2020,ke_pathophysiology_2022}. However, it remains unknown whether the use of coarse race\footnote{Throughout the manuscript, we refer to \emph{race} groups for succinctness and consistency, but certain groups may be more accurately described as countries-of-origin or ethnicities, as we describe below. The difficulty in accurately describing this variable speaks to the messiness and complexity of such data as a whole.} categories conceals meaningful disparities in \emph{clinical risk score performance}: e.g., whether clinical risk scores perform differently for Indian or Chinese patients than for Asian patients as a whole. This question has been challenging to study in part due to lack of granular race data in widely used clinical machine learning datasets.

Here, we study this question by using a sample of 418K emergency department visits for which granular race data has recently been made available. We examine the performance of five predictive risk scores using four metrics of algorithmic performance, stratifying performance both across four coarse race groups (White, Black, Hispanic/Latino, and Asian) and 26 granular race groups. We find that nearly every predictive risk score and metric exhibits racial performance disparities that are obscured by only assessing performance at the coarse race group level. These disparities are not only statistically significant, but also practically significant: the variation in performance within coarse groups, at the granular level, often exceeds the variation in performance between coarse groups. For example, performance varies \emph{more} across the five granular groups coarsely coded as ``Asian'' than it does across the Asian, Black, White, and Hispanic/Latino coarse groups for multiple outcomes and metrics of performance. In other words, the granular racial variation concealed by using coarse categories can exceed the coarse racial variation that has been the focus of an enormous amount of work on algorithmic fairness in medicine~\citep{chen_why_2018,obermeyer_dissecting_2019,zink_race_2023,boulware_systemic_2021,seyyed-kalantari_underdiagnosis_2021,adam_write_2022}. 
 
We examine why these disparities emerge, in terms of properties of the underlying data distributions. We find that granular race groups vary significantly in their presenting symptoms $X$, in their rates of outcomes $y$, and in the relationship between the symptoms $X$ and the outcomes $y$. In other words, every critical aspect of the data distribution $p(X, y)$ varies significantly across granular race groups. (As we discuss, these differences likely arise due to many factors, including social determinants of health, since race groups are a social construct and map imperfectly onto biological concepts like genetic ancestry \citep{cerdena_race-based_2020,borrell_race_2021,oni-orisan_embracing_2021,ioannidis_recalibrating_2021,roberts_abolish_2021}.) These distributional findings imply that, beyond the specific risk scores we examine, other risk scores may also exhibit significant granular performance disparities. 

\subsection*{Generalizable Insights about Machine Learning in the Context of Healthcare}

Our findings have implications for both healthcare dataset providers and machine learning researchers. The disparities we observe imply that healthcare dataset providers should record and release granular, self-identified patient race whenever possible, as recent large clinical databases have done \citep{johnson_mimic-iv_2023, all_of_us_research_program_investigators_all_2019}. For researchers studying disparities in clinical machine learning, we show that it is important to examine algorithmic disparities by granular race, because the use of coarse race can hide significant \granularvariation. In instances where granular race data is not available, our results suggest caution when interpreting racial disparities in performance: in particular, even if performance does not appear to vary at the coarse group level, it may still vary at the granular group level, and studies at the coarse level may understate the true racial variation.

%% file: mlhc-camera-ready-template/02_related_work.tex
\section{Related work}

Most work on quantifying racial disparities in healthcare in the United States relies on the Census categories: White, Black, Asian, Hispanic/Latino\footnote{In the United States, ``Hispanic/Latino'' is an \textit{ethnicity} which can overlap with multiple race groups. For the purposes of this study, we refer to it as a coarse \textit{race} group, since (1) our dataset does not have separate race and ethnicity fields (the MIMIC data warehouse only includes ``race"), and (2) analogous to the other coarse groups, the Hispanic group is composed of many granular identities.}, Hawaiian/Pacific Islander, and Native American \citep{hanna_towards_2020}. Two lines of work relate  most closely to our own: critiques of widely-used race categories and studies of the substantial heterogeneity within coarse groups.

\paragraph{Critiques of Race Categories.} The Census categories have been criticized for their coarseness \citep{AHRQ_race_2018,kauh_critical_2021,borrell_race_2021,shimkhada_capturing_2021,lett_conceptualizing_2022}, lack of clear definitions \citep{tehranian_whitewashed_2008, omi_racial_2014, christian_global_2019}, and U.S.-centrism \citep{roth_methodological_2017,hanna_towards_2020}. Many works suggest the adoption of new taxonomies. \cite{denton_racial_1997}, \cite{saperstein_capturing_2012}, and \cite{roth_multiple_2016} advocate for explicit distinction between self-identified race and perceived race. \cite{bailey_measures_2013} demonstrate how switching between different measurements of race can have significant effects on the magnitude of estimated racial disparities in income. \cite{howell_so_2017} argue that salient race categories in sociological research should capture observed inequalities in income, housing, and health, and they propose a modification to the Census categories accordingly. Our work contributes to this growing body of literature by examining the impact of the coarse race taxonomy on studies of algorithmic fairness in health.

\paragraph{Studies of Granular Variation.} Prior work has shown that coarse race categories conceal meaningful heterogeneity in demographic features, including household income, education, and healthcare access \citep{mccracken_cancer_2007,torres_stone_beyond_2007,dorsey_heterogeneity_2017,read_disaggregating_2021}. 
These differences have led many in the health disparities community to call for more granular race variables \citep{anderson_racial_2004,wang_information_2020,flanagin_updated_2021,lett_conceptualizing_2022,caggiano_health_2022}, and led to studies that characterize heterogeneity in measures of health and well-being within racial subgroups \citep{mccracken_cancer_2007,dorsey_heterogeneity_2017,read_disaggregating_2021,nyc_health_2022}.
\cite{lett_conceptualizing_2022} highlight this phenomenon in the case of the Hispanic/Latino category, where its coarseness erases important ``cultural, linguistic, and racial diversity in Latin America."
Guatemalans and Cubans, for example, differ significantly in terms of both average income and immigration status.
\cite{caggiano_health_2022} use genetic data to detect descent-based granular groups and then use these groups to quantify disparities in healthcare utilization, clinical diagnoses, and genetic predispositions. Researchers in algorithmic fairness have similarly argued that fairness analyses involving a single, coarse demographic attribute are ethically and practically insufficient \citep{hanna_towards_2020, wang_towards_2022}.
Our work advances these literatures with a thorough empirical audit of granular heterogeneity in predictive performance.
While finer-grained race data has been studied in the context of specific health conditions (for example, chronic kidney disease \citep{kataoka-yahiro_asian_2019} and disability \citep{read_disaggregating_2021}), there have been no studies of how performance of clinical risk scores differs across granular groups.

In this work, we focus on 
 studying racial disparities in clinical risk score performance. A related but distinct topic is whether race corrections---i.e., including race as a predictive feature---should be included to ameliorate racial disparities. This topic has a rich body of related work \citep{vyas_hidden_2020,cerdena_race-based_2020,borrell_race_2021,oni-orisan_embracing_2021,ioannidis_recalibrating_2021,roberts_abolish_2021}, but is outside the scope of this paper.

%% file: mlhc-camera-ready-template/03_methods.tex
\section{Methods}
\label{sec:methods_disparities}

We analyze racial variation in the performance of clinical risk scores using multiple prediction tasks and performance metrics. Our analysis relies on the fact that each patient in our dataset has both a self-identified \emph{coarse race group} (e.g.~``Asian'') and a \emph{granular race group} (e.g.~``Indian'' or ``Chinese''), with granular race groups nesting within coarse race groups. We measure risk score performance separately for each coarse race group and each granular race group. We assess whether there is statistically significant variation in risk score performance across the granular groups within each coarse group and compare the magnitude of the variation between coarse groups to the variation within coarse groups. In this section, we further describe our dataset---MIMIC-IV-ED \citep{johnson_mimic-iv-ed_2023}, a dataset of emergency department vists---and analysis. Code to reproduce our results is available at \url{https://github.com/rmovva/granular-race-disparities_MLHC23}.

\paragraph{Cohort \& race data} To study disparities, we use the patient self-reported race variable in MIMIC-IV-ED. 
The cohort consists of 418K\footnote{We filter out $\sim$7,000 ED visits (roughly 1\% of the overall dataset) with patient age $< 18$, or with no recorded ED triage severity.} emergency department (ED) visits by 201K distinct patients to the Beth Israel Deaconess Medical Center (BIDMC) in Boston, MA.
Patients select from coarse categories like ``Asian'' or more specific categories like ``Asian - Chinese.''
Table \ref{table:race-counts} provides the list of coarse groups, their granular subgroups, and the counts of unique patients \& ED stays in each group.
To determine the mapping from granular to coarse groups, we followed MIMIC's coding scheme and US Census guidelines.
Note that the majority of White and Black patients (92\% and 83\% respectively) and a minority of Hispanic and Asian patients (9\% and 39\% respectively) only reported a coarse race category, and did not report a more specific race category.
We include these patients in our analysis as their own granular group, and use an asterisk to denote them. 
For example, for the ``Asian'' coarse group, we analyze 5 granular groups: ``Chinese'', ``Indian'', ``Southeast Asian'', ``Korean'', and ``Asian*'', where the final group consists of all patients who report that they are Asian without reporting a more specific Asian subgroup. 
We verify that whether a patient reports a more specific subgroup does not vary depending upon structural factors (e.g., arrival year or insurance status; Figure \ref{fig:granular_race_availability}).
More details on race categories are provided in Appendix \ref{sec:methods_racecategories}.

\input{mlhc-camera-ready-template/table_race_counts}

\paragraph{Prediction tasks \& features} To measure algorithm performance, we focus on predicting three emergency department outcomes. 
We largely follow \citet{xie_benchmarking_2022}: using their code, we extract 64 features, including age, sex, nurse-determined triage severity scores, vitals at triage, patient history (comorbidities; number of recent hospital and ICU visits), and chief patient complaints (full list in Table \ref{table:feature-names}).
We quantify performance on the same three clinical tasks as \citet{xie_benchmarking_2022}, each of which has been widely studied in ED medicine: 
(1) \textbf{hospitalization}: at triage, identifying patients who will be hospitalized ($\sim$45\% of ED visits); 
(2) \textbf{critical outcomes}: at triage, identifying patients who will experience inpatient mortality or an ICU transfer in 12h ($\sim$6\% of visits); 
(3) \textbf{revisit}: at discharge, identifying patients who return to the ED within 72h ($\sim$3\% of visits).
These outcomes are the subject of much prior work in emergency medicine and all relate to providing efficient, well-tailored patient care \citep{sun_predicting_2011, hong_predicting_2018, churpek_derivation_2012, martin-gill_risk_2004, pellerin_predicting_2018}; more context is provided in Appendix \ref{sec:methods_tasks}.

\paragraph{ED risk prediction models} We assess disparities in two types of scores: (1) previously-developed, clinically-studied scoring rules and (2) machine learning models that are trained on the MIMIC-ED dataset. 
For existing clinical scores, we study two measures for patient triage: the National Early Warning Score (NEWS; \citet{smith_ability_2013}) and the more-specialized Cardiac Arrest Risk Triage (CART; \citet{churpek_derivation_2012}). 
These scores are simple linear functions of vital signs and age, and are designed to identify the most at-risk patients who visit the ED. 
Since the scores may have been developed and studied in non-representative patient samples, we are interested in studying \granularvariation\ in their predictive utilities. 
For ML risk scoring, we train logistic regressions (LR) for each of the three outcomes. 
We use the same protocol as \citet{xie_benchmarking_2022} and verify with cross-validation that our model performance matches the metrics they report.
We also replicated results with XGBoost decision trees to ensure that our results are unaffected by model complexity, finding that the predictions were indeed highly concordant (Spearman $\rho \ge 0.85$; Table \ref{table:xgb}). Further details are in Appendix \ref{sec:methods_riskscores}.

\paragraph{Performance metrics} Past work in algorithmic fairness has used numerous metrics to evaluate whether algorithms perform equally well across groups~\citep{kleinberg_inherent_2016,chouldechova_fair_2017,narayanan_translation_2018,corbett-davies_measure_2018,chen_ethical_2021,zink_race_2023,mitchell_algorithmic_2021,corbett-davies_algorithmic_2017}. 
These metrics often conflict: one cannot simultaneously equalize all metrics across groups except in restrictive special cases~\citep{kleinberg_inherent_2016,chouldechova_fair_2017}. 
The proper choice of metric is context-specific and depends on the decision the algorithm is designed to inform~\citep{chen_ethical_2021}. 
Given this, and because we evaluate multiple models and prediction tasks, we measure performance using four common metrics: area under the precision-recall curve (AUPRC); area under the receiver-operating characteristic curve (AUROC); false positive rate (FPR); and false negative rate (FNR)\footnote{In addition, we also assess calibration error for the trained ML risk scores, revealing similar results to the other four metrics. See Appendix \ref{sec:supp_calibration} for details.}. 
FPR and FNR are computed using the thresholds given in \citet{xie_benchmarking_2022}. 
These metrics are widely used in the ML and algorithmic fairness literature, and using multiple metrics allows us to assess whether the performance disparities we observe emerge robustly regardless of the particular metric chosen.

\paragraph{Uncertainty quantification}
\label{para:uncertainty}

Our estimates of algorithmic performance across \granulargroup s will naturally vary due simply to statistical noise, particularly for smaller \granulargroup s, even in the absence of true differences in performance. Our goal is to quantify whether the variation in estimated performance we observe exceeds that expected due to noise, making it imperative to properly quantify uncertainty. We summarize our procedure for doing so here and provide full details in Appendix \ref{sec:appendix_quantifying_uncertainty}. To quantify uncertainty in the performance of the machine learning methods, we report the 95\% confidence interval across 1,000 random train-test splits, a widely used procedure~\citep{chen_intimate_2021,shanmugam_quantifying_2022}. We estimate uncertainty for the predefined risk scores (NEWS and CART), which do not require a train set, via a 95\% confidence interval across 1,000 bootstrapped datasets. This is a standard procedure for quantifying uncertainty~\citep{efron_introduction_1994} and is widely used in medical applications~\citep{mihaylova_review_2011,myers_identifying_2020,kompa_second_2021}. Throughout the manuscript, we sometimes perform many comparisons simultaneously --- for example, comparing each granular group to the corresponding coarse group across all outcomes and performance metrics. We provide specific details below, but note that whenever we perform such analyses, we perform Bonferroni multiple hypothesis correction \citep{dunn_multiple_1961} on all $p$-values, as is standard.

\paragraph{Mathematical notation} Following previous work, we let $\features$ denote the features for each patient, $\riskscore = f(\features)$ the risk score, and $\outcome \in \{0, 1\}$ the ground truth outcome. We use $\Agranular$ to denote the patient's \granulargroup~and $\Acoarse$ to denote their \coarsegroup. 

%% file: mlhc-camera-ready-template/table_race_counts.tex
% \usepackage{booktabs}
% \usepackage{multirow}
\begin{table}[!ht]
\small
\centering
\caption{\label{table:race-counts} \textbf{Coarse-to-granular group mapping as collected in the MIMIC database.} Counts of unique patients and ED stays for each group are listed. The asterisk * denotes patients who only reported a coarse race: e.g., ``Asian*'' indicates patients who self-identified as Asian and did not provide a more specific category. Overall, $\sim$20\% of patients report a granular race group that is distinct from their coarse group. ``SE Asian'': Southeast Asian. $N$ is the total number of patients per coarse group.}

\begin{tabular}{@{}llrr@{}}
\toprule
Coarse                                                                              & Granular           & Patients  & Stays \\ \midrule
\multirow{5}{*}{\begin{tabular}[c]{@{}l@{}}Asian\\ $N = $ 11K \end{tabular}}                                                              & Asian*             & 4,997   & 7,215     \\
                                                                                    & Chinese            & 4,027   & 7,271     \\
                                                                                    & Indian             & 859    & 1,549     \\
                                                                                    & SE Asian           & 828    & 1,512     \\ 
                                                                                    & Korean             & 500    & 774      \\ \midrule
\multirow{4}{*}{\begin{tabular}[c]{@{}l@{}}Black\\ $N = $ 32K \end{tabular}}                      & Black*             & 25,496  & 76,118    \\
                                                                                    & Cape Verdean       & 2,677   & 7,588     \\
                                                                                    & African            & 2,349   & 4,837     \\
                                                                                    & Caribbean          & 1,574   & 3,625     \\ \midrule
\multirow{6}{*}{\begin{tabular}[c]{@{}l@{}}White\\ $N = $ 126K \end{tabular}}                                                              & White*             & 117,403 & 224,969   \\
                                                                                    & Other Eur.          & 4,221   & 8,916     \\  
                                                                                    & Russian            & 2,041   & 6,018     \\          
                                                                                    & Brazilian          & 820    & 1,466     \\
                                                                                    & Eastern Eur.        & 611    & 1,297     \\
                                                                                    & Portuguese         & 586    & 1,427     \\ \midrule
\multirow{11}{*}{\begin{tabular}[c]{@{}l@{}}Hispanic/ \\Latino\\ $N = $ 14K \end{tabular}}      & Hispanic/Latino* & 2,019   & 3,070     \\
                                                                                    & Puerto Rican       & 4,169   & 13,913    \\
                                                                                    & Dominican          & 3,060   & 8,260     \\
                                                                                    & Guatemalan         & 991    & 2,323     \\
                                                                                    & Mexican            & 671    & 1,252     \\
                                                                                    & Salvadoran         & 633    & 1,482     \\
                                                                                    & Colombian          & 595    & 1,296     \\                                                             
                                                                                    & South American     & 496    & 1,055     \\                                                    
                                                                                    & Honduran           & 357    & 995      \\                                                             
                                                                                    & Central American   & 306    & 780      \\
                                                                                    & Cuban              & 250    & 779      \\ \midrule
\end{tabular}

\end{table}

%% file: mlhc-camera-ready-template/04_describing_disparities.tex
\section{Quantifying model performance disparities across granular race groups}
\label{sec:quantifying_disparities}

\begin{figure}[!htb]
    \centering
    \includegraphics[width=\textwidth]{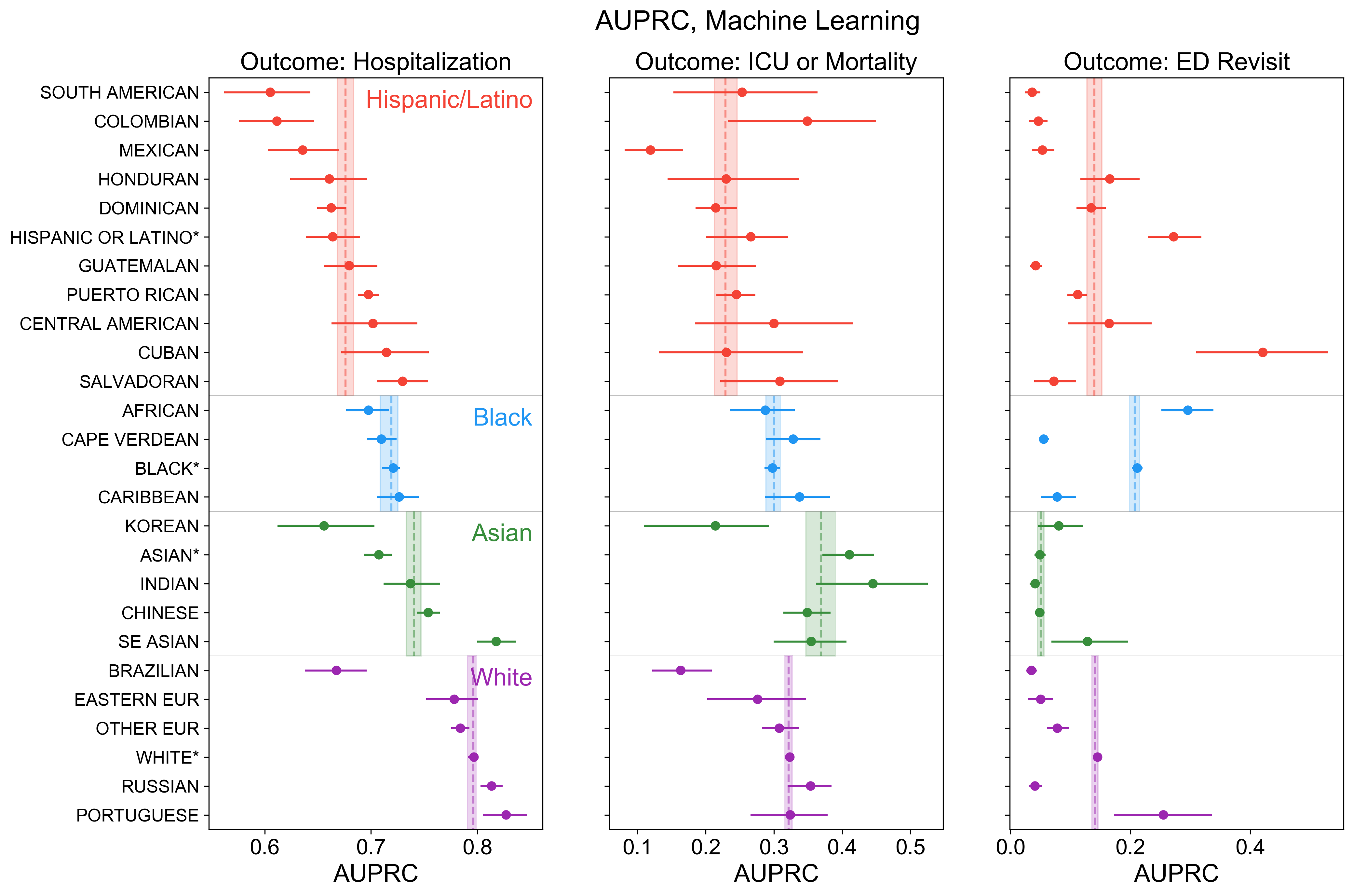}
    \caption{\textbf{Granular AUPRCs for machine learning risk scores trained on MIMIC-ED.} Points show medians and 95\% confidence intervals for granular group AUPRC across 1,000  runs. Dashed lines and shaded regions show medians \& CIs for coarse groups. Granular groups labeled with an asterisk * are the patients who only reported a coarse race.}
    \label{fig:lr_auprc}
\end{figure}

As described in Section \ref{sec:methods_disparities}, we evaluate the performance of five clinical risk scores (two previously developed scores and three machine learning models) in predicting three ED outcomes (hospitalization; ICU/mortality; and ED revisit). We assess disparities in model performance by computing AUPRC, AUROC, FPR, and FNR for each coarse group and each granular group, and assessing whether performance in each granular group differs significantly from performance in the corresponding coarse group after multiple hypothesis correction.

Figure \ref{fig:lr_auprc} plots AUPRC for the machine learning models, revealing that many granular groups exhibit performance which differs significantly from the performance of the overall coarse group. Examining predictive performance for hospitalization, for example (Figure \ref{fig:lr_auprc} left), reveals that model performance on patients who report their granular race group as South American, Colombian, or Mexican is worse than performance on Hispanic/Latino patients overall; conversely, model performance on Salvadoran patients is better. Within the White coarse group, Brazilian patients experience significantly worse risk score performance than the group as a whole for all three outcomes, while within the Asian coarse group, Koreans and Southeast Asians are often outliers. Figures \ref{fig:lr_auroc}-\ref{fig:lr_fnr} show analogous results for the other three metrics --- AUROC, FPR, and FNR --- revealing significant variation across the board. 
(In Appendix \ref{sec:supp_calibration} and Figure \ref{fig:lr_calibration}, we additionally show that calibration error varies significantly across granular groups.)

\input{mlhc-camera-ready-template/table_permutation_test_logreg}

Table \ref{table:logreg_pval} extends our analysis to all performance metrics and outcomes. For each outcome, metric, and coarse race group, we report whether performance on at least one granular race group within the coarse race group differs statistically significantly from overall coarse group performance, after multiple hypothesis correction for the number of tests performed. All metrics, outcomes, and coarse race groups exhibit at least one statistically significant disparity, demonstrating that examining performance at only the coarse group level consistently conceals important granular variation. Among metrics, AUPRC, FNR, and FPR exhibit disparities somewhat more consistently than AUROC; this may be driven in part by variation in base rates across granular groups, as we explore further in Section \ref{sec:explaining_differences}. Among outcomes, ``Hospitalization'' and ``ED revisit'' exhibit more consistent performance disparities across granular groups than does the ``Critical'' (ICU or mortality) outcome. 

The standard risk scores, NEWS and CART, also exhibit significant performance differences (Tables \ref{table:news} and \ref{table:cart} in the appendix). Compared to the machine learning models, NEWS and CART exhibit lower performance overall, but they nonetheless exhibit many of the same granular disparity trends. For the hospitalization and critical outcomes, for example, the Spearman correlation of granular AUPRCs between NEWS and the machine learning model was $\sim$0.9. NEWS and CART did not exhibit many performance disparities for the ED revisit outcome, since here they yield very poor performance across all groups.

\begin{figure}[!htb]
    \centering
    \includegraphics[width=\textwidth]{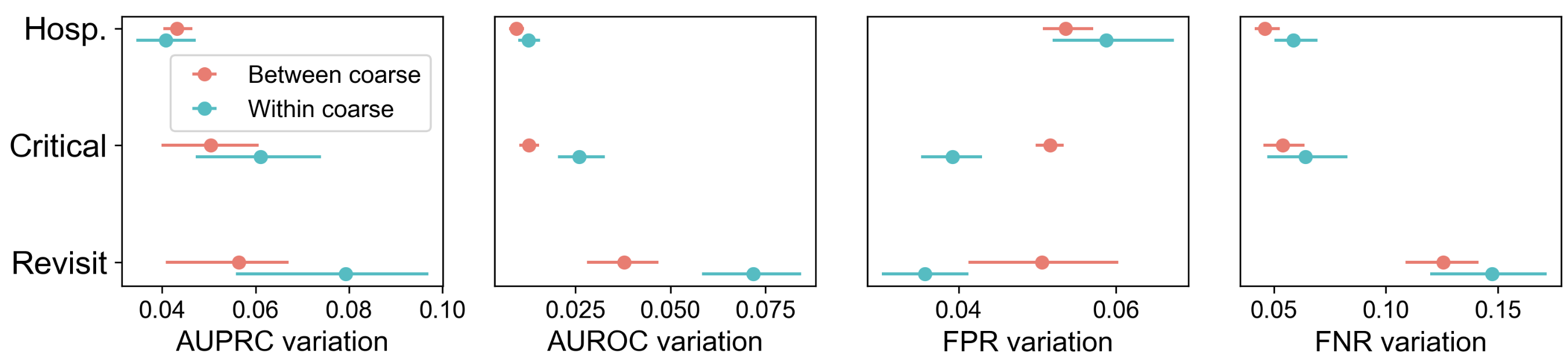}
    \caption{\textbf{Within-coarse-group variation is comparable or larger than between-coarse-group variation.} Here, variation is defined as the standard deviation of performance across the four coarse groups (\textit{between}) or the average SD across the granular groups \textit{within} each coarse group. Error bars are 95\% CIs across 1,000 train/test shuffles.}
    \label{fig:variation}
    \vspace{-.3cm}
\end{figure}

Having established the existence of statistically significant performance disparities within coarse groups, we compare the magnitude of \emph{within-coarse-group} variation (i.e., across granular groups within a coarse group) and \emph{between-coarse-group} variation. Between-coarse-group variation corresponds to what is assessed by many previous algorithmic fairness analyses of health datasets. We quantify between-coarse-group variation as the standard deviation in a performance metric across the four coarse groups. To quantify within-coarse variation, for each of the four coarse groups, we compute the standard deviation in performance across granular groups within that coarse group. We then take the unweighted average across the four coarse groups. Intuitively, these two measures compare the variation in performance across the four coarse race groups to the average variation in performance across granular groups within each coarse race group. In Figure \ref{fig:variation}, we plot 95\% CIs of these two variation measures across the 1,000 train/test shuffles.

We find that within-coarse-group variation (blue) is typically comparable to or larger than between-coarse-group variation (red). 
For 9 of 12 outcome/metric pairs, the within-coarse-group variation point estimates exceed the between-coarse-group estimates, and in some cases they are more than twice as large. 
These comparisons highlight the magnitude of the variation concealed by analyzing only coarse groups: the concealed variation is often larger than the between-coarse-group variation which has been the subject of study for the vast majority of previous work on fairness in clinical machine learning.\footnote{We confirmed that we did not merely observe this result because granular groups are generally smaller than coarse groups (that is, granular performance metrics are estimated using less data than coarse metrics, which might inflate estimates of within-coarse-group metric variation). Specifically, we recomputed our estimates of between-coarse-group variation after downsampling the coarse groups to be, on average, the same size as the granular groups (Fig.~\ref{fig:supp_variation}). 
% (We chose this factor so that the amount of data used to estimate between-coarse-group variation matched the amount of data used to estimate within-coarse-group variation for the average coarse group.) 
Though the between-coarse-group variation confidence intervals widened, as expected, the between-coarse-group variation point estimates were still smaller than the within-coarse-group variation point estimates for the same outcome/metric pairs as before.
}

%% file: mlhc-camera-ready-template/table_permutation_test_logreg.tex
\begin{table}
\centering

\caption{\textbf{Granular variation in performance of machine learning models trained on MIMIC-ED.} For each metric and coarse group, asterisks denote whether there is at least one granular group with significantly different predictive performance than the coarse group. All $p$-values are computed with Bonferroni multiple hypothesis correction. $\star$: $p < 0.05$, $\star\star$: $p < 0.01$, $\star\star\star$: $p < 0.001$, \texttt{-} not significant.}

% \textbf{Outcome} & \textbf{Metric} & \textbf{AUPRC} & \textbf{AUROC} &  \textbf{FNR} &  \textbf{FPR} \\

\label{table:logreg_pval}
\begin{tabular}{l|lcccc}
\toprule
    \textbf{Outcome}    & \textbf{Coarse} &              AUPRC &              AUROC &                FPR &                FNR \\
\midrule
\multirow{4}{*}{\textbf{Hospitalization}} & \textbf{Asian} &  $\star\star\star$ &                  - &  $\star\star\star$ &  $\star\star\star$ \\
        & \textbf{Black} &                  - &  $\star\star\star$ &  $\star\star\star$ &  $\star\star\star$ \\
        & \textbf{Hispanic/Latino} &       $\star\star$ &                  - &  $\star\star\star$ &  $\star\star\star$ \\
        & \textbf{White} &  $\star\star\star$ &                  - &  $\star\star\star$ &  $\star\star\star$ \\
\midrule
\multirow{4}{*}{\textbf{Critical}} & \textbf{Asian} &                  - &            $\star$ &  $\star\star\star$ &                  - \\
        & \textbf{Black} &                  - &       $\star\star$ &  $\star\star\star$ &                  - \\
        & \textbf{Hispanic/Latino} &       $\star\star$ &                  - &            $\star$ &                  - \\
        & \textbf{White} &  $\star\star\star$ &                  - &  $\star\star\star$ &       $\star\star$ \\
\midrule
\multirow{4}{*}{\textbf{Revisit}} & \textbf{Asian} &                  - &                  - &  $\star\star\star$ &                  - \\
        & \textbf{Black} &  $\star\star\star$ &  $\star\star\star$ &  $\star\star\star$ &  $\star\star\star$ \\
        & \textbf{Hispanic/Latino} &  $\star\star\star$ &       $\star\star$ &  $\star\star\star$ &  $\star\star\star$ \\
        & \textbf{White} &  $\star\star\star$ &  $\star\star\star$ &  $\star\star\star$ &  $\star\star\star$ \\
\bottomrule
\end{tabular}

\end{table}

%% file: mlhc-camera-ready-template/05_explaining_disparities.tex
\section{Explaining differences in algorithmic performance across granular race groups}
\label{sec:explaining_differences}

Thus far, we have established that significant granular variation in performance exists within each coarse group. We now explore why these disparities emerge by studying aspects of each granular group's underlying data distribution. In particular, we study the role of differences in sample sizes (\S \ref{sec:diff_in_sample_size}); outcome frequencies, $p(y)$ (\S \ref{sec:diff_p_y}); feature distributions, $p(X)$ (\S \ref{sec:diff_p_x}); and feature-outcome relationships, $p(y \mid X)$ (\S \ref{sec:diff_p_y_given_x}). We conduct this analysis for several reasons. First, it can deepen our understanding of why we observe the performance disparities documented in \S \ref{sec:quantifying_disparities}. Second, it informs whether we would expect to observe similar disparities in other risk scores (beyond those examined in \S \ref{sec:quantifying_disparities}): if many aspects of the data distribution differ across granular groups, we might expect to see other risk scores show disparities as well. Finally, depending on what aspects of the data distribution differ, there are different solutions to disparities in algorithmic performance: for example, if $p(X)$ differs across groups, one might improve performance by employing techniques designed to address covariate shift \citep{singh_fairness_2021}.

\subsection{Differences in sample size}
\label{sec:diff_in_sample_size}

We first ask whether differences in group sample sizes might explain the predictive disparities we observe in the machine learning models. 
Past work has shown that unequal training dataset representation can lead to worse performance for underrepresented groups \citep{chen_why_2018, buolamwini_gender_2018}, so it is natural to test this hypothesis given how some granular groups are much larger than others (Table \ref{table:race-counts}). 
Computing Spearman correlations between granular group size and performance, we find that for most metrics and outcomes, there is surprisingly no significant relationship. 
In particular, none of the four metrics are significantly correlated with group size for the hospitalization and critical outcomes.
For the revisit outcome, AUPRC and FPR are significantly correlated with group size.
The lack of correlation suggests that variation in the distributions of $X$ and $y$, rather than dataset representation, may better explain the disparities we observe.

\subsection{Differences in outcome frequency, $p(y)$}
\label{sec:diff_p_y}

\begin{figure}[!ht]
    \centering
    \includegraphics[width=\textwidth]{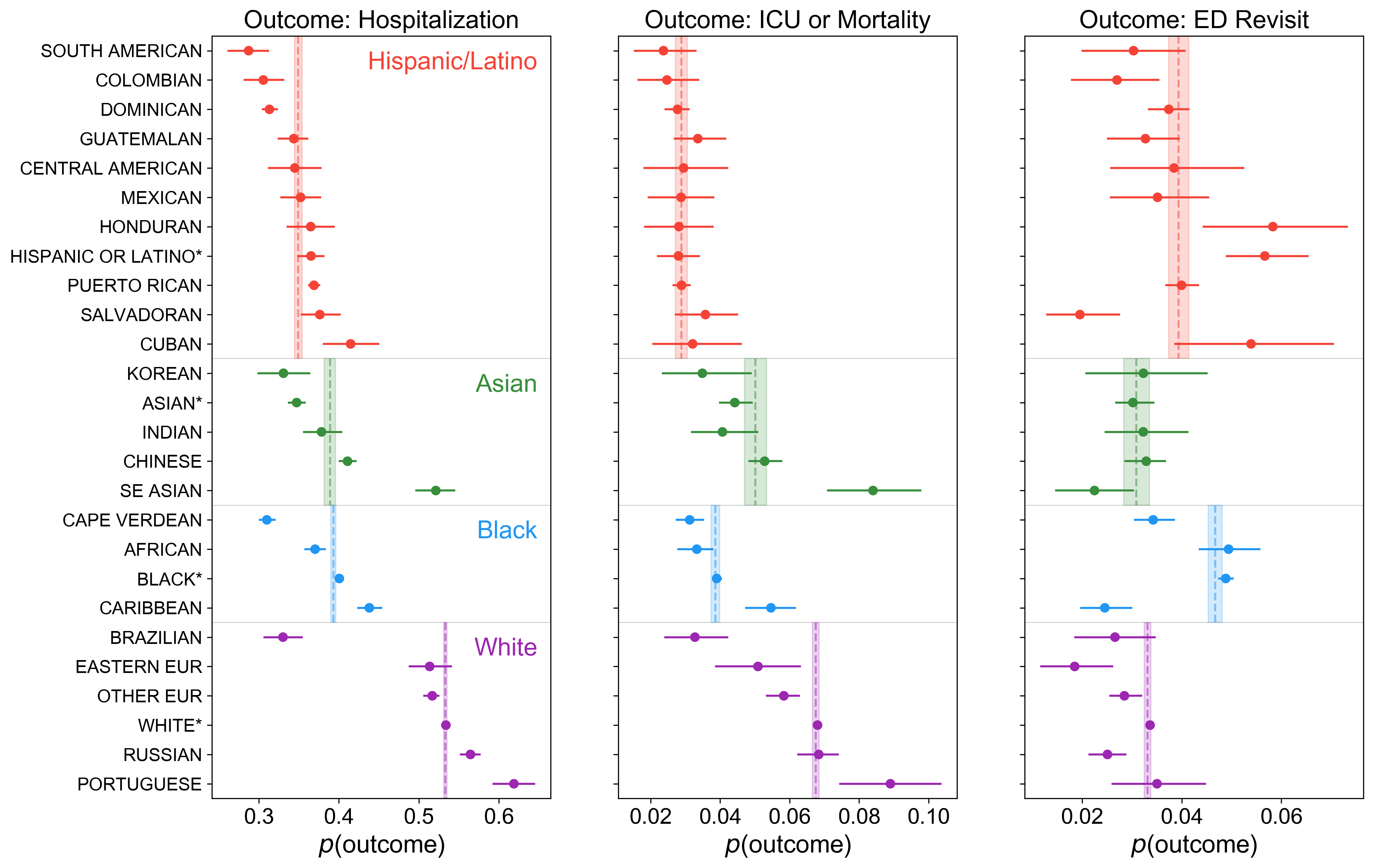}
    \caption{\textbf{Outcome frequencies differ by granular group.} Points show outcome frequencies with bootstrapped 95\% confidence intervals. Dashed lines and shaded regions show outcome frequencies \& CIs for coarse groups. *Granular groups labeled with an asterisk are the patients who only reported a coarse race.}
    \label{fig:outcome_freqs}
\end{figure}

Patient groups may differ in their underlying outcome rates, $p(y)$, and these differences can propagate to predictive metric differences for a trained model. 
For example, the baseline AUPRC for a random classifier is $p(y=1)$ \citep{saito_precision-recall_2015}, so AUPRC will naturally tend to be higher for groups with higher outcome frequency.
Figure \ref{fig:outcome_freqs} plots $p(y=1 \mid A^{(g)})$ for each outcome, revealing substantial and statistically significant granular variation in outcome frequency. 
Further, many of the previously mentioned groups with outlier performances---including Brazilians, Koreans, and SE Asians---are exactly the groups with outlier outcome frequencies. 
The granular predictive metrics also differ from the coarse predictive metrics in the direction we would expect given differences in outcome frequency: for example, since Brazilians are hospitalized less frequently than White patients as a whole, the model ends up under-predicting Brazilian patient risk, which causes more false negatives and fewer false positives. 
Across the 26 granular groups, granular AUPRC, FPR, and FNR display significant Spearman correlations with $p(y \mid A^{(g)})$ (Spearman $\rho$: 0.56--0.88). These results suggest that one reason we observe disparities across granular groups in AUPRC, FPR, and FNR is that the outcome frequencies differ across granular groups.

These results further underscore the importance of granular analyses: differing outcome rates are both important to study on their own, and can hint at causes of disparities in clinical risk score performance.
However, differences in $p(y \mid A^{(g)})$ do not fully explain why we observe the performance disparities in \S \ref{sec:quantifying_disparities}: consider that there is significant granular variation in AUROC (Table \ref{table:logreg_pval}), even though we observe no correlation between granular AUROC and $p(y \mid A^{(g)})$ (and there is no mathematical reason why they should correlate). 
Thus, there must be other sources of granular variation contributing to the disparities in risk score performance, which we explore next.

\subsection{Differences in feature distributions, $p(X)$}
\label{sec:diff_p_x}

We now examine the extent to which granular race groups differ in terms of their distribution over $X$ (i.e., covariate shift \citep{shimodaira_improving_2000} between granular race groups). We use two representations of patient symptoms: ICD codes and Elixhauser comorbidity index (ECI) codes, which are binary indicators of the presence of comorbidities \citep{elixhauser_comorbidity_1998}. Table \ref{tab:enriched_icd_codes} lists ICD codes that are significantly more common in a granular group, compared to the remainder of the coarse group. Table \ref{tab:enriched_eci_codes} replicates this analysis using ECI codes. Treating the Indian granular race group as an example, we estimate the prevalence of a particular code among Indian patients, and divide this number by the prevalence of that code among the \emph{remaining} Asian patients---that is, in reference to patients within the coarse group who do not identify as the granular group. We identify significantly enriched codes by using a Fisher exact test after applying Bonferroni multiple hypothesis correction. Specifically, we adjust for the 149,630 ICD code comparisons (26 granular groups $\cdot$ 5755 ICD codes) and 806 ECI code comparisons (26 granular groups $\cdot$ 31 ECI codes). We report up to five codes per granular group, sorted by magnitude of enrichment, and exclude ICD codes and ECI codes that occur fewer than 10 times in a particular granular group for privacy reasons. %We note that the reported trends may be specific to the hospital and region in which this data was collected. 

Substantial differences emerge, many of which are supported by prior literature. Within the Asian coarse group, enriched comorbidities include  hypothyroidism in Indians \citep{talwalkar_prevalence_2019}, kidney failure in Chinese patients \citep{liyanage_prevalence_2022}, and alcohol abuse in Korean patients \citep{yom_advancing_2022}. Enriched ICD codes in the Hispanic/Latino and White coarse groups---e.g., the higher prevalence of Hepatitis C among Puerto Rican patients \citep{perez_epidemiology_2013,nyc_health_2022} and heart failure in Russian patients \citep{townsend_cardiovascular_2016}---also align with existing literature. Black patients who report a more specific granular group (i.e., patients who self-identify as ``Black - Cape Verdean'', ``Black - African'', or ``Black - Caribbean'') have fewer comorbidities compared to Black patients who do not report a more specific group (recorded as ``Black*" in Tables \ref{tab:enriched_icd_codes} and \ref{tab:enriched_eci_codes}). One possible explanation for this is that Black patients who report a more specific group are more likely to be immigrants. Previous work has found that foreign-born Black patients experience a lower prevalence of cardiovascular disease, maternal health, and diabetes compared to their US-born counterparts \citep{collins_differing_2002,read_racial_2005,dorsey_heterogeneity_2017,turksonocran_comparison_2020}, a phenomenon that has been referred to as the ``healthy immigrant effect" \citep{antecol_unhealthy_2006}.

\subsection{Differences in feature-outcome relationships, $p(y \mid X)$}
\label{sec:diff_p_y_given_x}

Another source of predictive disparities could be that the feature-outcome relationships---the mappings from features $X$ to outcomes $y$---vary with $A^{(g)}$. 
The risk scores (both ML and clinical) assume that the presence of a feature has the same risk implications for all patients.

To test whether $p(y \mid X)$ depends on $A^{(g)}$, we compare two simple regression designs with and without granular race interaction terms. 
That is, for each set of patients in a given coarse group, we include granular race as a categorical covariate and compare the logistic regressions (\texttt{LR}):
\begin{align}
    \texttt{y} &\sim \texttt{LR(X, granular\char`_race)} \\
    \texttt{y} &\sim \texttt{LR(X, granular\char`_race, X*(granular\char`_race))}
\end{align}

Regression (1) includes only granular-race-specific offset terms, while Regression (2) also allows the coefficient for each feature to differ for each granular race group. If $p(y \mid X)$ varies with granular race within a given coarse group, we would expect that second regression explains statistically significantly more variation in $y$ than the first, adjusting for the fact that it has more parameters and thus more capacity to explain variation. 
To assess this, we use a likelihood ratio test \citep{vuong_likelihood_1989} to compare the goodness-of-fit of the two regressions across coarse groups and outcomes. For the hospitalization and critical outcomes, the likelihood ratio test strongly rejects the null ($p < 10^{-6}$) for all coarse groups, indicating that Regression (2), with granular-specific coefficients, better fits the data (Table \ref{table:supp_lrtest}).
This indicates that the feature-outcome relationships vary significantly within coarse race groups.

Next, we examine \emph{which} features exhibit different relationships by granular race group.
To do so, we modify Regression (2) to include granular race interaction terms one at a time, for each feature. That is, for each coarse group and feature \texttt{x\char`_i}, we run the regression 
\begin{align} \texttt{y} &\sim \texttt{LR(X, granular\char`_race, x\char`_i*(granular\char`_race))}, \tag{3} \end{align} and use an likelihood ratio test to compare to Regression (1). 
For a given coarse group, the resulting $p$-value tests whether that feature's association with the outcome varies with granular race.

The results of these tests for all pairs of features and coarse race groups are given in Tables \ref{table:feature_interactions_crit} and \ref{table:feature_interactions_hosp} for the critical and hospitalization outcomes, respectively; we only show the features/race pairs that are significant after Bonferroni correction.
For the critical outcome, there are two features with significant granular variation in $p(y \mid x_i)$ for the White coarse group, seven for Black, eight for Hispanic/Latino, and nine for Asian. 
Surprisingly, there are more features which show statistically significant granular variation for the non-White groups, even though they are smaller and thus have reduced statistical power.
One important feature whose weight varies across granular groups is triage acuity, which is an index from 1 to 5 assigned by ED nurses to categorize patient severity.
If acuity scores were assigned consistently based on risk of deterioration, we would expect the same acuity coefficient for all groups in predicting ICU transfer/mortality.
However, three of the four coarse groups display significant granular heterogeneity in the acuity coefficient, suggesting that the acuity measure may be more tailored to some groups than others. 
This finding aligns with prior work, which finds disparities in triage scores across coarse race groups \citep{schrader_racial_2013, boley_investigating_2022}. 
There are also several examples of comorbidity features with granular coefficient variation, implying that the same comorbidities have different predictive relationships with outcomes depending on granular race.
Again, such differences have been studied between coarse groups \citep{howard_racial_2013, spanakis_raceethnic_2013}, but we offer preliminary evidence that feature-outcome relationships are yet another component of our data distribution which display granular variation.

%% file: mlhc-camera-ready-template/06_discussion.tex
\section{Discussion}

We show that stratifying clinical risk score performance only by coarse race group can conceal significant disparities in performance across granular race groups. 
Our subsequent analysis of why these disparities arise finds that granular groups differ in terms of outcome rates $p(y)$, presenting symptoms $p(X)$, and the relationship between features and outcomes $p(y \mid X)$.  
Our results suggest that it is imperative for healthcare dataset providers to collect granular race data, and for researchers to stratify model performance by granular race group, not only by coarse group. 
Analyses stratified only by coarse race groups may overlook salient disparities in predictive performance.
While we document this pitfall for clinical risk scores, our findings also have implications for the many other settings where coarse categories have been used to study inequality and algorithmic bias~\citep{chetty2020race,goel_precinct_2016,franchi2023detecting,rho2023escalated,kleinberg2018human,voigt2017language,kline2022systemic,laufer2022end,pierson2020assessing,liu2022equity,derenoncourt2021minimum,chouldechova_fair_2017,garg_word_2018,cheng_marked_2023,bianchi_easily_2023, abdu_empirical_2023}, suggesting the importance of examining granular race categories in these domains as well. 

Our findings have limitations. 
First, our analysis only includes patients from a single ED. 
As a result, our cohort is specific to one region---Boston---and precludes any generalizations about specific granular groups in other geographies.
It is likely that granular disparities exhibit patterns that are both hospital- and region-specific \citep{baicker_geographic_2005}, so further work is necessary to explore how these disparities replicate across hospitals. 
A multi-ED analysis may observe larger racial disparities than we do, since past work finds variation across hospitals in algorithmic performance, and patient racial demographics can differ significantly by hospital \citep{lyons_factors_2023}.
Second, our analysis is specific to ED outcomes. 
Our findings on granular distribution shift suggest that results may generalize, though the specific effects likely depend on outcome.
We hope that future work extends our findings to other outcomes.
Third, our analysis relies on a particular mapping of granular to coarse race groups. 
While this mapping is certainly imperfect---one of the facts that motivates our analysis---the pervasive granular variation we find suggests that any mapping of granular groups to coarse groups is likely to obscure important disparities.

Our analysis studies (1) whether granular disparities in performance exist and (2) why these disparities arise. We leave the question of how to \emph{reduce} these disparities as a natural direction for future work, which dovetails with an enormous amount of research in algorithmic fairness~\citep{chen_ethical_2021,chen_why_2018,rezaei_robust_2021,shah_selective_2022}. The distributional differences we investigate in Section \ref{sec:explaining_differences} each suggest different solutions. For example, the existence of covariate shift (differences in $p(X)$) between granular race groups suggests that models trained on certain granular groups may not generalize to others \citep{nestor_feature_2019}, and that recent techniques to address covariate shift would be appropriate \citep{singh_fairness_2021}. A natural question, from a machine learning standpoint, might also be whether inclusion of granular race as a predictive feature would ameliorate the predictive disparities we observe, since the predictive risk scores we study, which are developed by previous work, do not include granular race as a feature. The inclusion of race as a predictive feature in clinical algorithms has been the subject of an enormous amount of research and debate \citep{vyas_hidden_2020,cerdena_race-based_2020,borrell_race_2021,oni-orisan_embracing_2021,ioannidis_recalibrating_2021,roberts_abolish_2021}, and this question lies beyond the scope of this work. We note that merely including an additive term in the fitted risk scores for each granular race group would not remove all the disparities we observe: for example, it would leave unchanged the AUROC and AUPRC for each granular group (since these metrics are invariant to monotone transformations), and thus the disparities in these metrics.

Race categories merit continual evaluation and re-evaluation for their ability to capture inequality in healthcare and in clinical machine learning. A number of interesting questions remain. Given the instability of self-identified race across time, place, and context \citep{saperstein_double-checking_2006,roth_multiple_2016}, how can we revise the process of granular race data collection to account for this uncertainty? From a methodological perspective, how do we design analyses that are robust to inconsistencies in self-identified race in existing datasets? How do we resolve differences in the meaning of race between countries, and move towards a global methodology for quantifying health disparities? Our work is one step towards the goal of better representing, and ultimately mitigating, algorithmic disparities in health.

%% file: mlhc-camera-ready-template/99_appendix.tex
\setcounter{figure}{0}
\makeatletter 
\renewcommand{\thefigure}{S\@arabic\c@figure}
\makeatother

\setcounter{table}{0}
\makeatletter 
\renewcommand{\thetable}{S\@arabic\c@table}
\makeatother

\section{Supplementary Methods}
\label{appendix:methods}

\subsection{Additional context for race categories} 
\label{sec:methods_racecategories}

\begin{figure*}
    \centering
    \includegraphics[width=\columnwidth]{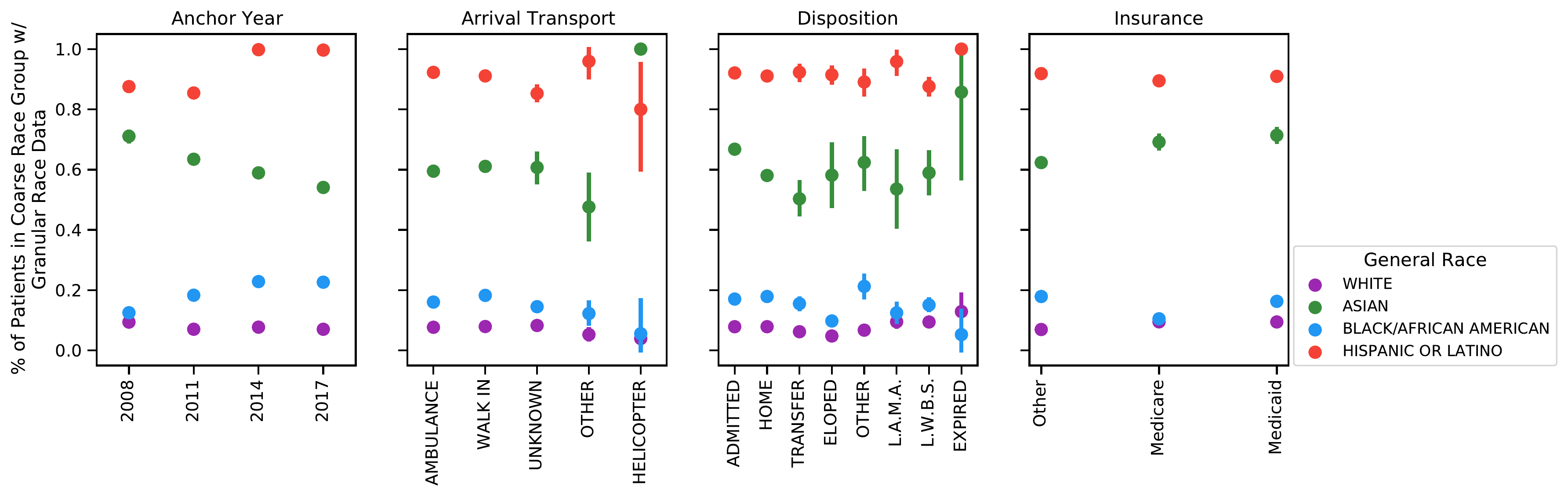}
    \caption{\textbf{The availability of granular race data demonstrates no clear relationship with structural factors.} We examine the dependence of granular race availability on the approximate year the patient appeared in the emergency department (anchor year), arrival transport, disposition, and insurance status to determine whether race data collection may depend upon observable features. For each variable, there is no clear distinction between patients with coarse race data and patients with granular race data.}
    \label{fig:granular_race_availability}
\end{figure*}

We contacted the dataset authors to learn more about race/ethnicity data collection in MIMIC. The authors shared that race data is primarily collected during patient registration, and were not aware of any changes over time in how these data were collected. We verify that whether a patient reported granular race does not substantially correlate with structural factors, such as the year of the ED visit or how the patient was transported to the ED (Figure \ref{fig:granular_race_availability}). There are many possible reasons why some patients did not report a \granulargroup: e.g., (a) they do not identify with a more specific group; (b) they do identify with a granular group, but it wasn't listed on the form; (c) they do identify with a more granular group, but they did not know to report it. We cannot distinguish between these causes, but we hypothesize that the reason may vary by \coarsegroup.

The granular-to-coarse mapping we use (Table \ref{table:race-counts}) comes directly from the MIMIC data, in which most of the granular races were labelled as a coarse label followed by a granular label, e.g., ``Asian - Chinese'' or ``Hispanic/Latino - Colombian''. 
The two exceptions were Portuguese and South American, which we labelled as White and Hispanic/Latino respectively, following Census guidelines.
It's worth noting that this system of assigning granular identities to coarse groups is imperfect: for example, in our dataset, we follow the Census guideline in classifying Brazilian and Portuguese Americans as White and not Hispanic.
However, this decision is contested by some \citep{marrow_be_2003,lopez_who_2022}.
Ambiguities like this one capture a core issue with coarse groupings, where it can be unclear who to include under a broad group label. 
Granular races allow more of the population to be clearly made visible, rather than obscured by vague boundaries. 

\subsection{List of features}
\label{sec:methods_features}

Table \ref{table:feature-names} lists the 64 features we use to train ML-based risk scores for each outcome.
We borrow these features directly from \citet{xie_benchmarking_2022}.
The Charlson and Elixhauser comorbidity features are binary features, combining related ICD codes into a single indicator of whether a patient has a particular condition \citep{sharma_comparing_2021}.
For the vital sign features, values that were clearly invalid were removed and imputed to median values. 
The median was computed only on the training set to avoid test set contamination. 
The ranges for valid values were taken from the MIMIC-Extract paper, as is standard for ML studies on MIMIC \citep{wang_mimic-extract_2020}.

\subsection{Additional context on the studied ED outcomes} 
\label{sec:methods_tasks}

The three ED outcomes we study are (1) hospitalization, (2) critical cases, i.e., ICU transfer in 12h or in-hospital mortality, and (3) ED revisits within 72h after discharge. We used the code from \citet{xie_benchmarking_2022} to extract the labels for all three outcomes from the MIMIC-ED database. These prediction tasks all relate to providing rapid, well-tailored patient care and running the ED efficiently; they are widely studied as a result. Here, we cover more past work on each of these outcomes.

Predicting hospitalization (Outcome 1) can improve real-time hospital management via accurate estimates of ED-to-inpatient flow; past literature has proposed several models \citep{sun_predicting_2011, peck_predicting_2012, hong_predicting_2018}. Similarly, accurate predictions of patient deterioration (Outcome 2) can forecast ICU load and help allocate limited resources like hospital and ICU beds. Several early warning scores have been developed to identify deteriorating patients \citep{prytherch_viewstowards_2010, churpek_derivation_2012, alam_impact_2014}, with recent ML-based approaches \citep{muralitharan_machine_2021, romero-brufau_using_2021}, and emerging evidence suggests that warning systems reduce overall inpatient mortality \citep{escobar_automated_2020}. Finally, patients who revisit the ED within 3 days of discharge (Outcome 3) may have received inadequate care \citep{keith_emergency_1989}, and revisit rates are a common (but controversial) quality-of-care statistic for hospitals \citep{martin-gill_risk_2004, pham_seventy-two-hour_2011, trivedy_unscheduled_2015}. There has been past research on predicting revisits \citep{hayward_predictors_2018, pellerin_predicting_2018}, both to understand why they happen and whether they can be intervened on. 

In each task, performance variation across groups has important implications, both for patient care and for understanding quality-of-care \citep{seyyed-kalantari_underdiagnosis_2021}. The purpose of our paper is to assess whether the coarse race data currently available in most healthcare settings is sufficient to capture racial variation in predictive performance. Beyond the implications of our findings in the ED, we also believe that the chosen tasks are representative of the rich patient diversity in most clinical settings: patients from all demographic groups visit the ED, and to the extent that we observe disparities in ED risk prediction, there is potential for disparities in other clinical prediction tasks as well.

\subsection{Additional details on risk scores \& ML modeling}
\label{sec:methods_riskscores}

There were many possible clinical risk scores to study: NEWS, CART, NEWS2, MEWS \citep{subbe_validation_2001}, and REMS \citep{olsson_rapid_2004}, for example. 
After computing these scores, and looking at their correlation matrix across patients, we found that NEWS and CART captured the two primary clusters of variation; other scores were strongly correlated (Spearman $\rho > 0.7$) to either NEWS or CART, so we focused our analysis to those two.

For ML models, we train L2-regularized logistic regressions (LR). The LR models were trained using regularization strength \texttt{C=1.0}, chosen using grid search with cross-validation. Our model's performance metrics match the ranges reported by \citet{xie_benchmarking_2022}.

We didn't use more complex models because they do not provide significantly better predictive performance on these tasks as noted both by  \citet{xie_benchmarking_2022} and \citet{hong_predicting_2018} in a different hospital system. 
We also confirm this by replicating our experiments with XGBoost decision trees. We find that performance is within the confidence interval of the logistic regression.
Further, XGBoost displays strongly concordant performance trends across granular groups, so the disparity analysis is nearly identical.
Across granular groups, XGBoost has $\rho \ge 0.85$ Spearman correlation in performance with the logistic regression performance, for all metrics and outcomes.

\begin{figure}[!htb]
    \centering
    \includegraphics[width=0.7\textwidth]{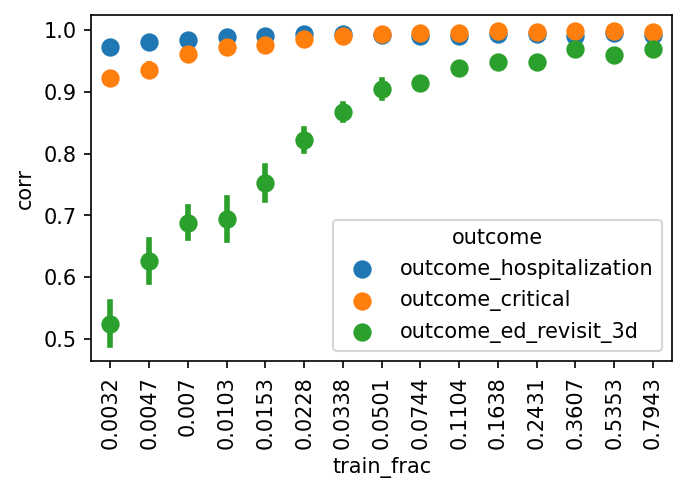}
    \caption{\textbf{Spearman correlation in hold-out predictions between a model trained on $x\%$ of the data vs. a model trained with 80\% of the data.} Note the log-scale of the $x$-axis. The model only needed about 30\% of the training data to achieve predictions that were nearly indistinguishable from the predictions of a model trained on 80\% of the data. As a result, we used a 30\%/70\% train/test split for our 1,000 shuffled experiments, because a larger test set gave us higher statistical power to detect performance disparities on the test set.}
    \label{fig:spearmancorr}
\end{figure}

We find that the logistic regressions do not need much training data to achieve maximal performance. Using only $\sim$30\% of the dataset, cross-validation AUC reaches a maximum for all outcomes, and predictions had near perfect Spearman correlation with the predictions from a model trained on 80\% of the data (Figure \ref{fig:spearmancorr}). Using a larger test set allows for higher-precision estimates of model performance (i.e., tighter confidence intervals on model test performance), which is especially important to allow for precise comparisons of performance between small granular subgroups. Therefore, we used a 30\%/70\% train-test split for all experiments in this paper. We split the dataset at the patient level, not the visit level, as is standard to prevent data leakage \citep{luo_guidelines_2016,tampu_inflation_2022}; thus, no patient appears in both the train and test set. \citet{xie_benchmarking_2022} do not split by patient, which may explain small performance discrepancies between our paper and theirs.

\subsection{Assessing calibration error of ML risk scores}
\label{sec:supp_calibration}

To supplement four commonly-studied metrics presented in the main text, we also study calibration of the ML risk scores. 
Calibration assesses how well predicted risk probabilities match the true probability of an outcome, and it is widely studied in ML and healthcare \citep{ crowson_assessing_2016, kleinberg_inherent_2016, nixon_measuring_2020, yadlowsky_calibration_2019, deniffel_using_2020, khurshid_ecg-based_2022}.
For example, a calibrated classifier would output a risk score of 0.8 for a patient with an 80\% risk of hospitalization.
Here, we assess calibration using binned expected calibration error (ECE) defined in \citet{pakdaman_naeini_obtaining_2015}; we use 10 bins (though we checked that the results are highly similar with other bin counts).
In this metric, predicted risks are binned into 10 deciles. In each decile $Q_m$, we compute the absolute difference between the average predicted risk $\hat{y}$, and the true proportion of patients with a positive label $y$, and then take an average of these differences:
$$\text{ECE}_{10\text{-bin}} = \frac{1}{10} \sum_{m=1}^{10} \text{abs}\left(\mathbb{E}_{i\in Q_m}[\hat{y}_i - y_i]\right).$$

\begin{figure}[!ht]
    \centering
    \includegraphics[width=\textwidth]{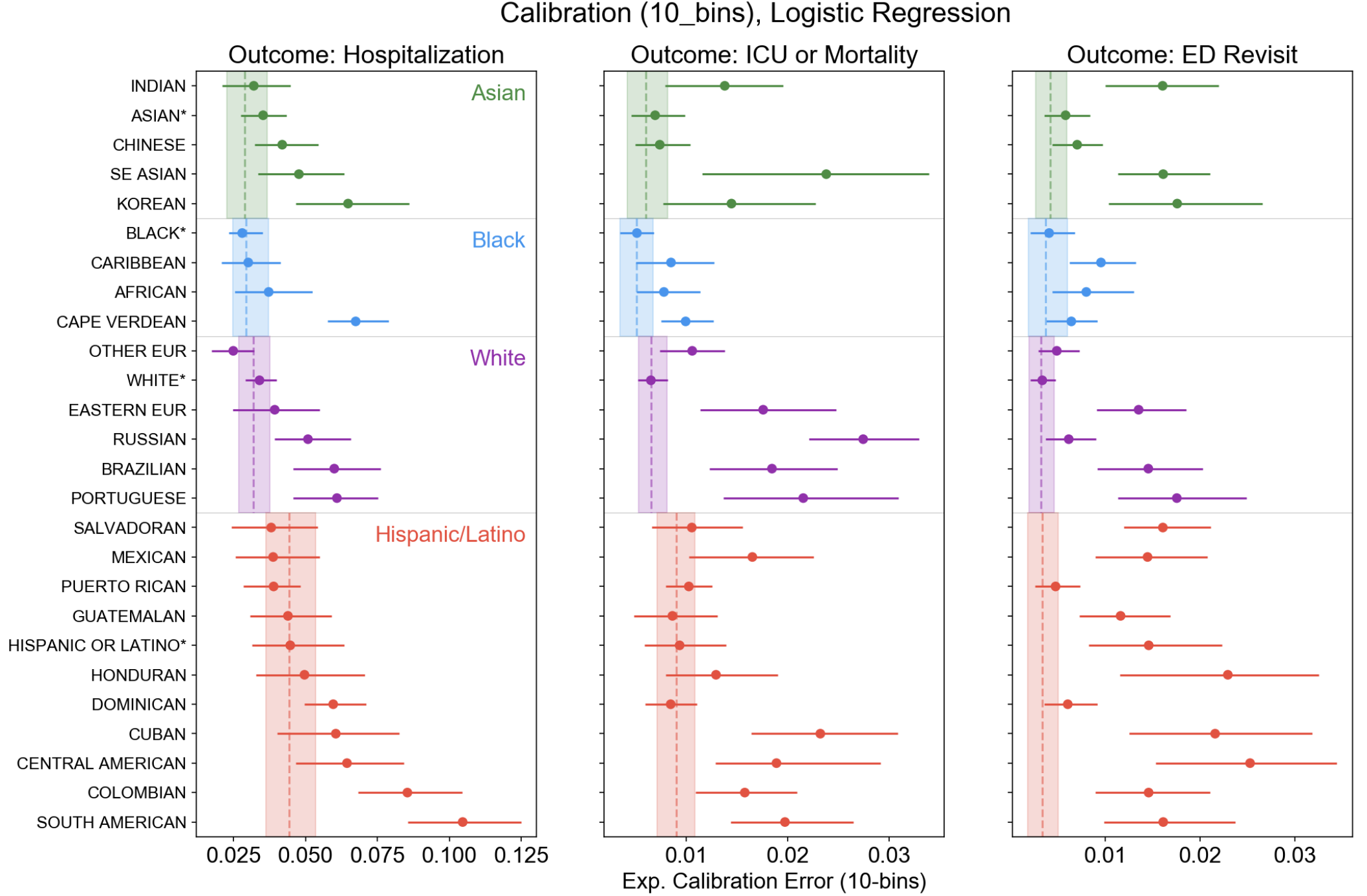}
    \caption{\textbf{Calibration error for machine learning risk scores trained on MIMIC-ED.} While the risk scores are relatively well-calibrated on the entire dataset, they are miscalibrated for certain groups. In particular, certain granular groups experience particularly poor calibration. See Appendix \ref{sec:supp_calibration} for more details on the definition of expected calibration error (ECE).}
    \label{fig:lr_calibration}
\end{figure}

We compute this calibration metric for the ML classifier for each of the three outcomes. 
We look at 10-bin ECE over the entire dataset, over coarse groups, and over granular groups. 
Over the entire dataset, the ML risk scores are well-calibrated for all tasks, with an ECE of 2.5\% for the hospitalization task and less than 0.3\% for the critical and revisit tasks.
The classifier is not as well-calibrated for certain groups; we show these results in Figure \ref{fig:lr_calibration}.
Specifically, calibration is similar for most coarse groups, but some granular groups are notable outliers. 
Using the same approach as in Table \ref{table:logreg_pval}, we find that for 8 of the 12 (coarse group, outcome) pairs, at least one granular group has a significantly different ECE than the coarse group average (with MH correction).
We conclude that calibration is yet another quality of risk scores which may vary with granular race: despite low calibration error over all patients and over coarse groups, the trained risk scores are significantly miscalibrated for some granular groups.

\subsection{Additional details on quantifying uncertainty}
\label{sec:appendix_quantifying_uncertainty}

To quantify uncertainty in machine learning performance, we employ a procedure widely used in previous work \citep{zink_race_2023,chen_intimate_2021,shanmugam_quantifying_2022}: we run 1,000 iterations, reshuffling the dataset each time; for each iteration, we randomly split the dataset into a train and test set, refit the model on the train set, and compute performance metrics on the test set. We report the 95\% confidence interval across shuffles (i.e., the 2.5th and 97.5th percentiles across the 1,000 shuffles). 

To quantify uncertainty in the performance of the predefined risk scores (NEWS and CART), we do not need a train set (since the procedure for computing the scores is defined by earlier work). Instead, we use bootstrapping, a standard procedure for quantifying uncertainty~\citep{efron_introduction_1994} which is widely used in medical applications~\citep{mihaylova_review_2011,myers_identifying_2020,kompa_second_2021}: for each iteration, we sample datapoints with replacement from the original dataset to produce a ``bootstrapped'' dataset of the same size as the original dataset; recompute performance metrics on the bootstrapped dataset; and repeat this procedure for 1,000 iterations. We report the 95\% confidence interval across bootstraps. 

We assess whether performance on a granular race group differs significantly from performance on the corresponding coarse group. 
To do so, we compute the $z$-score of the 1,000 differences in granular and coarse performance (i.e., mean divided by standard deviation over these 1,000 instances). 
We compute a two-tailed normal $p$-value for the $z$-score (we verify that normal distributions fit the data well, and note that normality assumptions are standard in many hypothesis tests). Because we examine many combinations of risk scores, outcomes, and coarse race groups, we perform Bonferroni multiple hypothesis correction on all $p$-values, multiplying them by 312 ($3 \text{ outcomes} \cdot 4 \text{ metrics} \cdot 26 \text{ granular groups} = 312 \text{ comparisons}$). 

\section{Supplementary Tables/Figures}
\label{appendix:results}

\input{mlhc-camera-ready-template/table_feature_names}

\input{mlhc-camera-ready-template/table_enriched_symptoms_fisher}
\input{mlhc-camera-ready-template/table_enriched_eci_codes_fisher}

% \subsection*{Analysis of granular race availability over observable features}

% \subsection*{Granular variation plots for additional metrics: AUROC, FPR, FNR}

\begin{figure}[!ht]
    \centering
    \includegraphics[width=\textwidth]{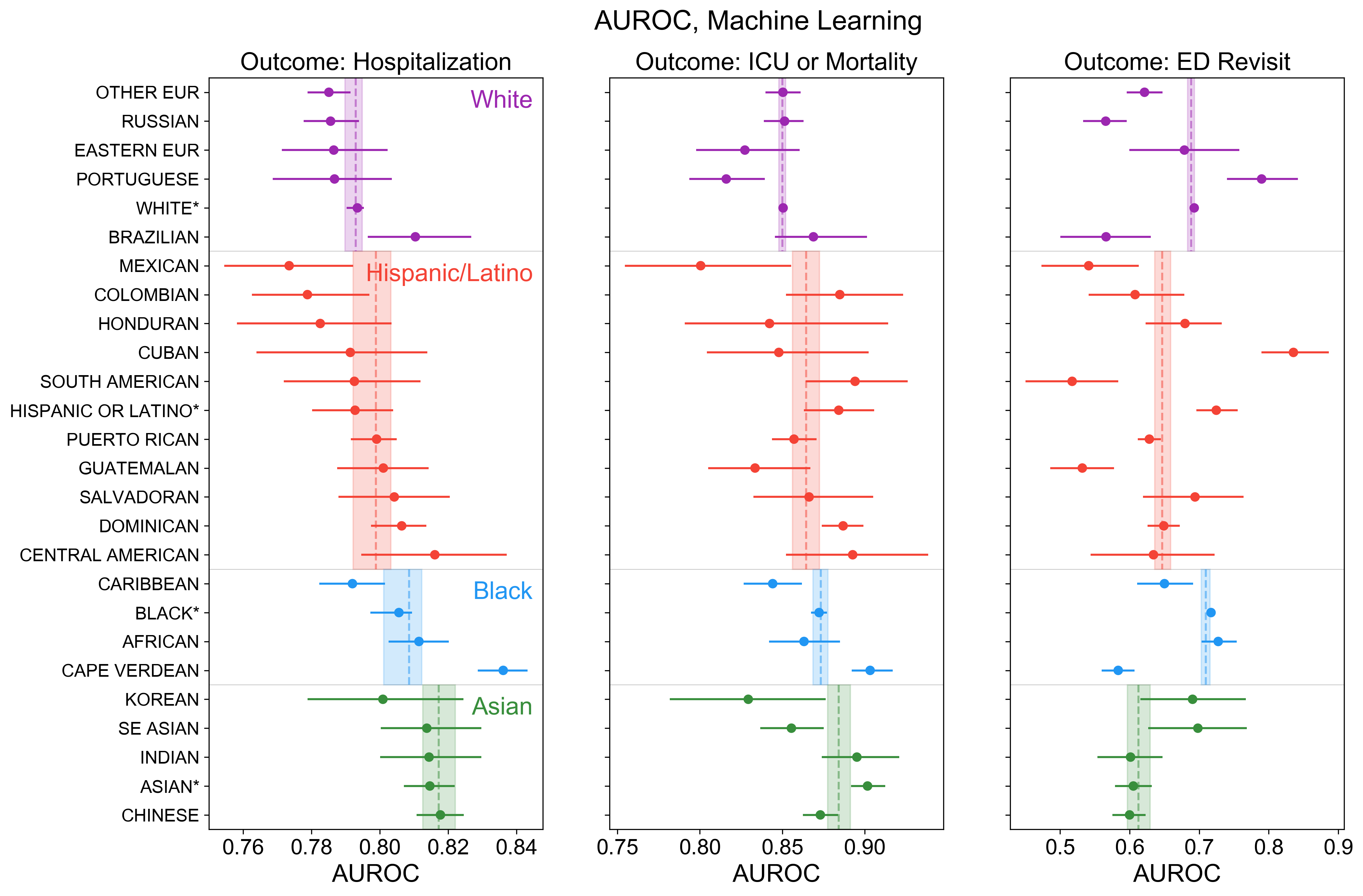}
    \caption{\textbf{Granular AUROCs for machine learning models (logistic regression) trained on MIMIC-ED.} Analogous to Figure \ref{fig:lr_auprc}.}
    \label{fig:lr_auroc}
\end{figure}

\begin{figure}[!ht]
    \centering
    \includegraphics[width=\textwidth]{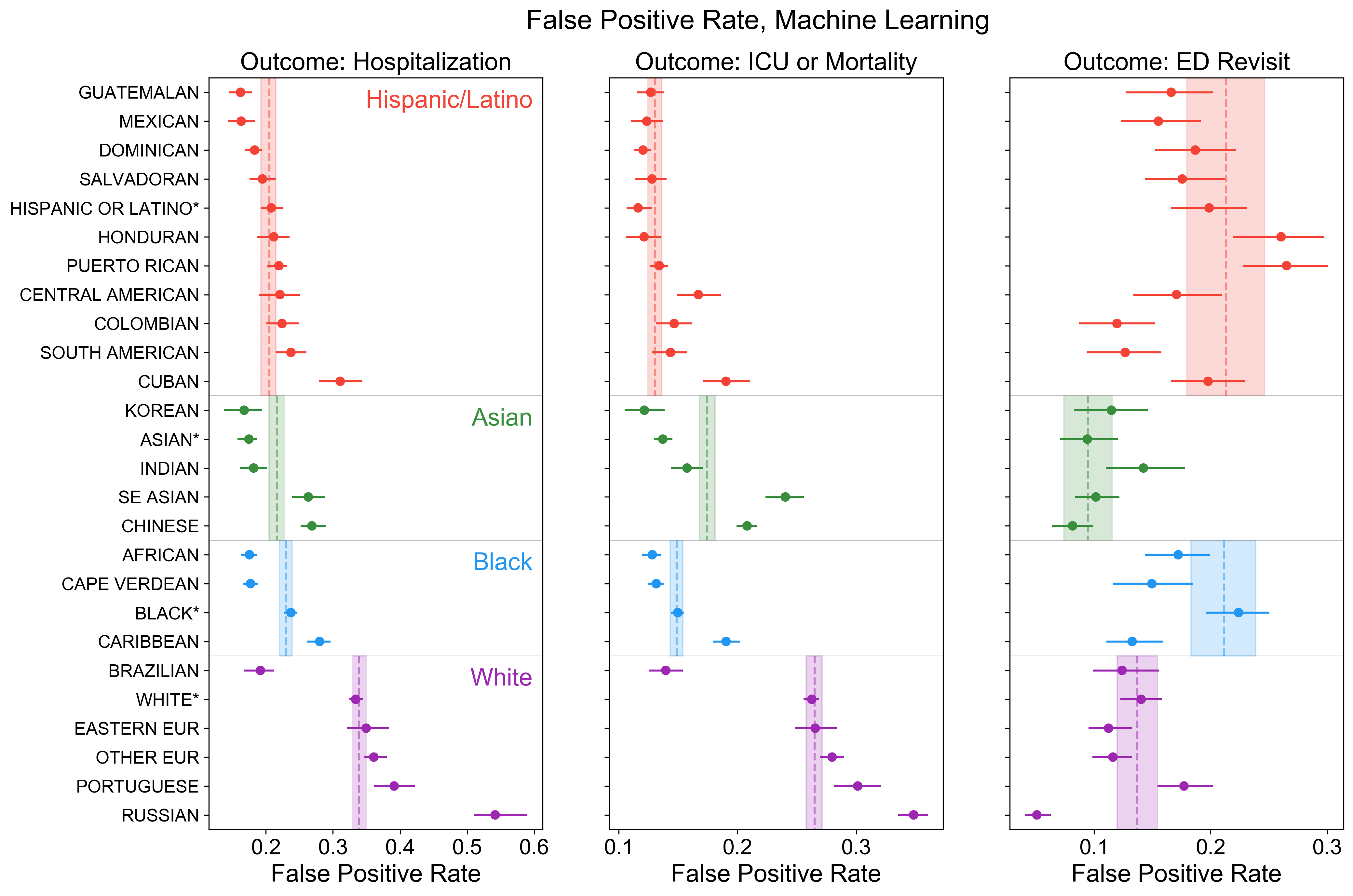}
    \caption{\textbf{Granular false positive rates (FPRs) for machine learning models (logistic regression) trained on MIMIC-ED.} Analogous to Figure \ref{fig:lr_auprc}.}
    \label{fig:lr_fpr}
\end{figure}

\begin{figure}[!ht]
    \centering
    \includegraphics[width=\textwidth]{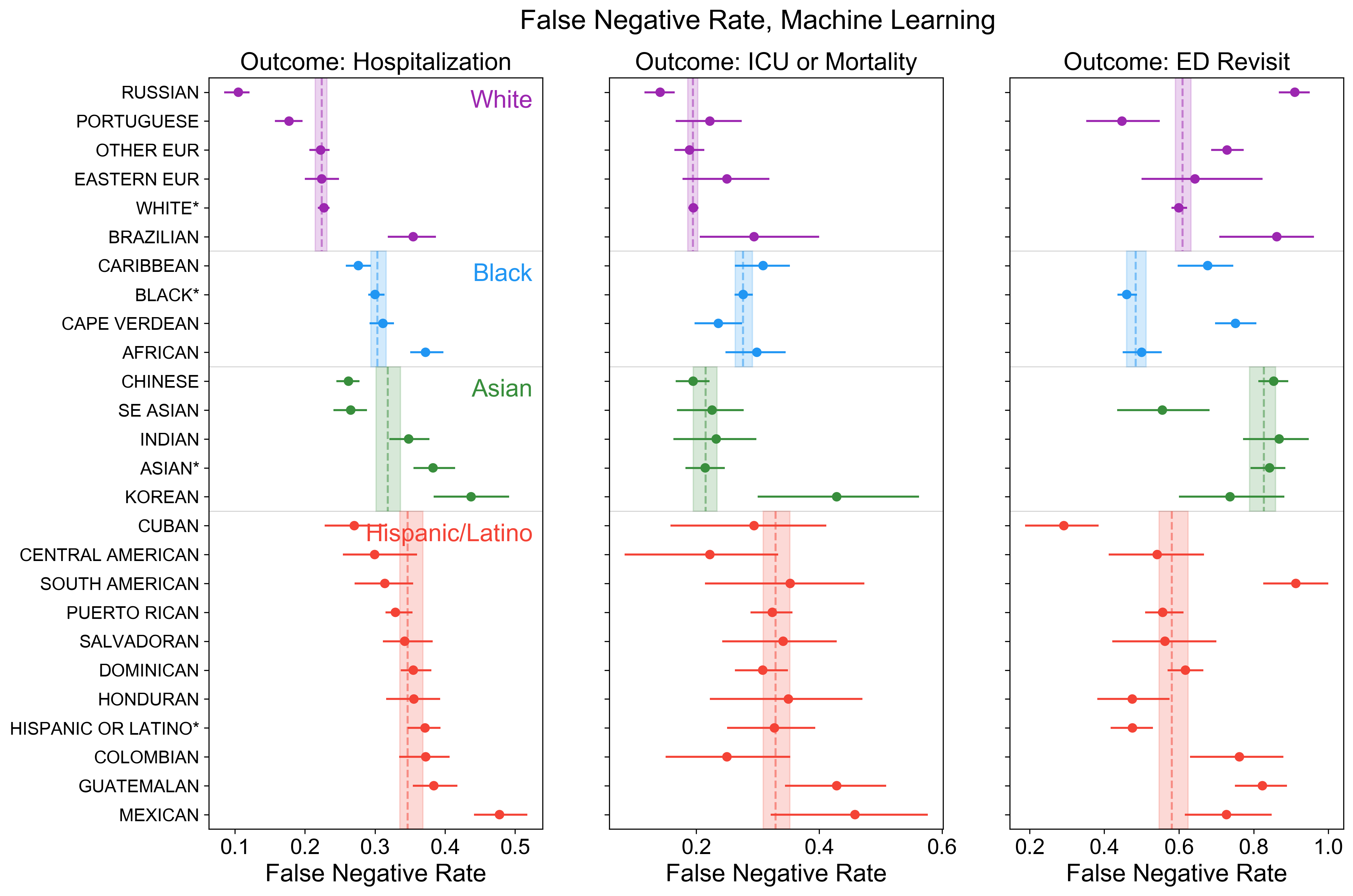}
    \caption{\textbf{Granular false negative rates (FNRs) for machine learning models (logistic regression) trained on MIMIC-ED.} Analogous to Figure \ref{fig:lr_auprc}.}
    \label{fig:lr_fnr}
\end{figure}

\begin{figure}[!ht]
    \centering
    \includegraphics[width=\textwidth]{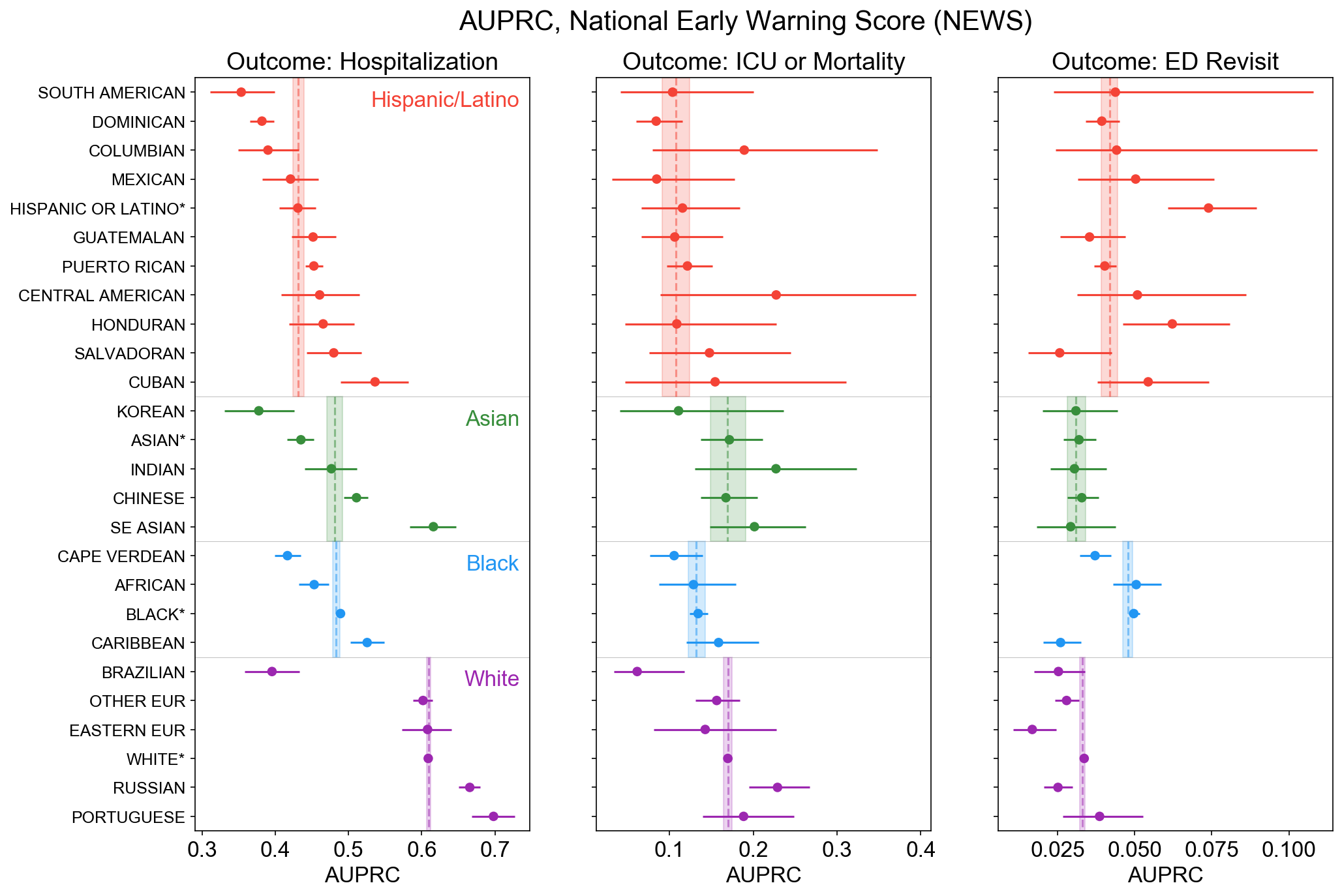}
    \caption{\textbf{Granular AUPRCs for the National Early Warning Score (NEWS), a previously-defined clinical risk score.} Analogous to Figure \ref{fig:lr_auprc}. The granular disparity trends for NEWS are similar to the ML models: compared to the ML models, the Spearman correlations for the median AUPRCs across granular groups are 0.93, 0.89, and 0.56 for the three outcomes, respectively.}
    \label{fig:news_auprc}
\end{figure}

% \subsection*{Tests for granular variation within coarse groups for clinical risk scores}

\input{mlhc-camera-ready-template/supp_pval_tables}

% \subsection*{Downsampled between- vs. within-coarse variation}

\begin{figure}[!htb]
    \centering
    \includegraphics[width=\textwidth]{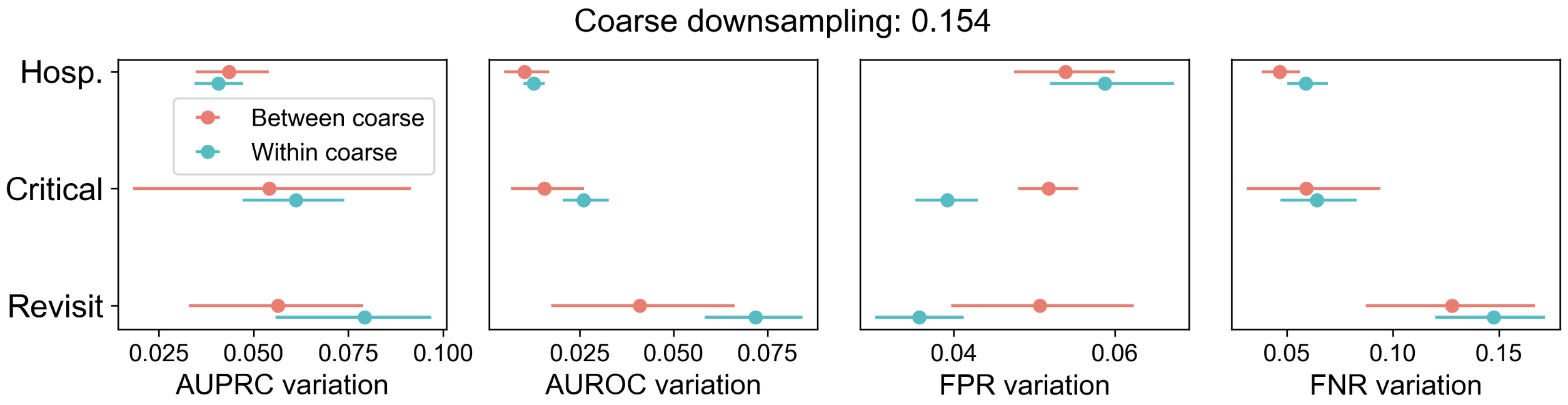}
    \caption{\textbf{After downsampling, within-coarse variation remains comparable to between-coarse variation.} We replicate this result after downsampling the coarse group sizes at a ratio of $\frac{N \text{ coarse groups}}{N \text{ granular groups}}$ ($\sim$0.15$\times$ their original size), so that the average number of patients in the downsampled coarse groups is equal to the average number of patients in a granular group.
    The between-coarse CIs widen, but the result that within-coarse variation is often larger than between-coarse variation still holds.
    Thus, this finding does not seem to be attributable to differences in coarse and granular group size.
    }
    \label{fig:supp_variation}
\end{figure}

% \subsection*{Table \ref{table:supp_lrtest}: likelihood-ratio tests for model interacted with granular race}

\begin{table}[]
\centering
\caption{\textbf{Likelihood ratio test $p$-values for a regression with interaction terms.} The regressions being compared are \texttt{y $\sim$ X + granular\_race} and \texttt{y $\sim$ X + granular\_race + X*(granular\_race)}, subsetting the data to one coarse race group at a time. The null hypothesis is that, when accounting for the additional parameters of the more complex model, the two regressions have the same goodness-of-fit. All $p$-values strongly reject the null, except for the Revisit outcome for the Asian coarse group. This means that the interaction terms improve the model's fit, i.e. the feature-outcome relationships $p(y \mid X)$ in our dataset vary with granular race.}
\label{table:supp_lrtest}
\begin{tabular}{@{}llll@{}}
\toprule
                & Hospitalization     & Critical            & Revisit             \\ \midrule
White           & $2 \times 10^{-12}$ & $2 \times 10^{-18}$ & $6 \times 10^{-5}$  \\
Black           & $3 \times 10^{-14}$ & $4 \times 10^{-26}$ & $1 \times 10^{-4}$  \\
Hispanic/Latino & $1 \times 10^{-28}$ & $2 \times 10^{-31}$ & $5 \times 10^{-25}$ \\
Asian           & $3 \times 10^{-7}$  & $4 \times 10^{-29}$ & 0.13                \\ \bottomrule
\end{tabular}
\end{table}

% \subsection*{Table \ref{table:feature_interactions_crit}, \ref{table:feature_interactions_hosp}: likelihood-ratio tests for individual feature interaction terms}

\input{mlhc-camera-ready-template/table_feature_interactions}

\input{mlhc-camera-ready-template/table_outcomes_scores_metrics}

%% file: mlhc-camera-ready-template/table_feature_names.tex
\begin{table}
    \centering
    \tiny
    
\caption{\textbf{Features used to train ML-based clinical risk scores.} We use 64 features, describing information on demographic group, visit frequency, chief complaint, and comorbidities, to train each of the ML-based clinical risk scores. The rightmost column contains observed ranges for each variable.}
\label{table:feature-names}

\begin{tabular}{llcc}
\toprule
                Category &                          Name &       Type &        Range \\
\midrule
             Demographic &                           Age & Continuous &     (18,103) \\
              &               Sex (True=Male) &     Binary & \{True,False\} \\
         Visit Frequency &     \# of ED visits within 30d & Continuous &       (0,20) \\
          &     \# of ED visits within 90d & Continuous &       (0,41) \\
          &    \# of ED visits within 365d & Continuous &      (0,112) \\
          &   \# of HOSP visits within 30d & Continuous &       (0,15) \\
          &   \# of HOSP visits within 90d & Continuous &       (0,30) \\
          &  \# of HOSP visits within 365d & Continuous &       (0,70) \\
          &    \# of ICU visits within 30d & Continuous &        (0,4) \\
          &    \# of ICU visits within 90d & Continuous &        (0,7) \\
          &   \# of ICU visits within 365d & Continuous &       (0,14) \\
                  Triage &               Temperature (C) & Continuous & (26.0,44.11) \\
                   &                     Heartrate & Continuous &  (1.0,256.0) \\
                   &              Respiratory Rate & Continuous &  (0.0,209.0) \\
                   &         Oxygen saturation (\%) & Continuous &  (0.0,100.0) \\
                   &       Systolic Blood pressure & Continuous &  (1.0,312.0) \\
                   &      Diastolic Blood Pressure & Continuous &  (0.0,375.0) \\
                   &                          Pain & Continuous &   (0.0,10.0) \\
                   &      Emergency Severity Index & Continuous &    (1.0,5.0) \\
         Chief Complaint &                    Chest pain &     Binary & \{True,False\} \\
          &                Abdominal pain &     Binary & \{True,False\} \\
          &                      Headache &     Binary & \{True,False\} \\
          &           Shortness of breath &     Binary & \{True,False\} \\
          &                     Back pain &     Binary & \{True,False\} \\
          &                         Cough &     Binary & \{True,False\} \\
          &               Nausea vomiting &     Binary & \{True,False\} \\
          &                  Fever chills &     Binary & \{True,False\} \\
          &                       Syncope &     Binary & \{True,False\} \\
          &                     Dizziness &     Binary & \{True,False\} \\
  Charlson Comorbidities &         Myocardial Infarction &     Binary & \{True,False\} \\
   &      Congestive Heart Failure &     Binary & \{True,False\} \\
   &      Peripheral Vasc. Disease &     Binary & \{True,False\} \\
   &       Cerebrovascular Disease &     Binary & \{True,False\} \\
   &                      Dementia &     Binary & \{True,False\} \\
   &         Chronic Pulm. Disease &     Binary & \{True,False\} \\
   &             Rheumatic Disease &     Binary & \{True,False\} \\
   &          Peptic Ulcer Disease &     Binary & \{True,False\} \\
   &            Mild Liver Disease &     Binary & \{True,False\} \\
   &     Diabetes W/o Complication &     Binary & \{True,False\} \\
   &      Diabetes W/ Complication &     Binary & \{True,False\} \\
   &                     Paralysis &     Binary & \{True,False\} \\
   &                 Renal Disease &     Binary & \{True,False\} \\
   &                    Malignancy &     Binary & \{True,False\} \\
   & Moderate/severe Liver Disease &     Binary & \{True,False\} \\
   &       Tumor, Metastatic Solid &     Binary & \{True,False\} \\
   &                      Aids/hiv &     Binary & \{True,False\} \\
Elixhauser Comorbidities &           Cardiac Arrhythmias &     Binary & \{True,False\} \\
 &              Valvular Disease &     Binary & \{True,False\} \\
 &     Pulmonary Circ. Disorders &     Binary & \{True,False\} \\
 &          Hypertension, Compl. &     Binary & \{True,False\} \\
 &        Hypertension, Uncompl. &     Binary & \{True,False\} \\
 &        Other Neuro. Disorders &     Binary & \{True,False\} \\
 &                Hypothyroidism &     Binary & \{True,False\} \\
 &                      Lymphoma &     Binary & \{True,False\} \\
 &                  Coagulopathy &     Binary & \{True,False\} \\
 &                       Obesity &     Binary & \{True,False\} \\
 &                   Weight Loss &     Binary & \{True,False\} \\
 & Fluid \& Electrolyte Disorders &     Binary & \{True,False\} \\
 &             Blood Loss Anemia &     Binary & \{True,False\} \\
 &             Deficiency Anemia &     Binary & \{True,False\} \\
 &                 Alcohol Abuse &     Binary & \{True,False\} \\
 &                    Drug Abuse &     Binary & \{True,False\} \\
 &                     Psychoses &     Binary & \{True,False\} \\
 &                    Depression &     Binary & \{True,False\} \\
\bottomrule
\end{tabular}
\end{table}

%% file: mlhc-camera-ready-template/table_enriched_symptoms_fisher.tex
\begin{table}[]
    \centering
    \tiny
\begin{tabular}{llll}
\toprule
 Coarse Race &       Granular Race &                                                     ICD Code &  Ratio \\
\midrule
       Asian &              Asian* &                 Alcohol abuse with intoxication, unspecified &   2.28 \\ \cmidrule{2-4} 
        &         Indian &                                  Hypothyroidism, unspecified &   2.88 \\
        &         &                          Unspecified acquired hypothyroidism &   2.85 \\ \cmidrule{2-4} 
        &             Chinese & Chronic viral hepatitis B without mention of hepatic coma... &   2.49 \\
        &              &                           Unspecified essential hypertension &   1.59 \\
        &              &                            Acute kidney failure, unspecified &   1.48 \\
        &              &                             Essential (primary) hypertension &   1.45 \\
        &              &                                          Anemia, unspecified &   1.41 \\ \cmidrule{2-4} 
        &              Korean &                                   Alcohol abuse, unspecified &   3.49 \\ \cmidrule{2-4} 
        &    SE Asian  &                            Acute kidney failure, unspecified &   1.85 \\ \midrule 
       Black &    Black* &         Other, mixed, or unspecified drug abuse, unspecified &   6.42 \\
        &     &                             Body Mass Index 45.0-49.9, adult &   5.56 \\
        &     &            Other psychoactive substance abuse, uncomplicated &   4.61 \\
        &     &                                                  Sarcoidosis &   4.61 \\
        &     &                       Body mass index (BMI) 50.0-59.9, adult &   4.31 \\ \cmidrule{2-4} 
        &        Cape Verdean & Other viral diseases in the mother, delivered, with or wi... &   4.31 \\
        &         &                                          Post-term pregnancy &   3.58 \\
        &         & Post term pregnancy, delivered, with or without mention o... &   3.05 \\
        &         & Second-degree perineal laceration, delivered, with or wit... &   2.57 \\
        &         &        Streptococcus B carrier state complicating childbirth &   2.26 \\ \cmidrule{2-4} 
        &    Caribbean Island & Nonspecific reaction to tuberculin skin test without acti... &   3.48 \\ \midrule 
Hisp./Latino & Hisp./Latino* & Suicide and self-inflicted injury by cutting and piercing... &   9.53 \\
 &  & Unspecified drug or medicinal substance causing adverse e... &   7.02 \\
 &  &       Acute alcoholic intoxication in alcoholism, continuous &   6.81 \\
 &  &                                             Other acute pain &   6.67 \\
 &  & Diabetes mellitus without mention of complication, type I... &   5.55 \\ \cmidrule{2-4} 
 &           Dominican & Other specified pregnancy related conditions, first trime... &   2.96 \\
 &            & Abnormality in fetal heart rate and rhythm complicating l... &   2.83 \\
 &            &                                            Single live birth &   2.09 \\
 &            &                             Essential (primary) hypertension &   1.37 \\ \cmidrule{2-4} 
 &        Puerto Rican & Poisoning by heroin, accidental (unintentional), initial ... &  12.27 \\
 &         &                                  Opioid abuse, uncomplicated &   4.46 \\
 &         &           Unspecified viral hepatitis C without hepatic coma &   3.06 \\
 &         &               Nicotine dependence, cigarettes, uncomplicated &   2.47 \\
 &         &                            Unspecified asthma, uncomplicated &   1.96 \\ \midrule 
       White &          Portuguese &                                          Portal hypertension &   3.90 \\ \cmidrule{2-4} 
        &              White* &       Acute alcoholic intoxication in alcoholism, continuous &   9.01 \\
        &               & Driver of heavy transport vehicle injured in collision wi... &   5.77 \\
        &               &                           Unspecified episodic mood disorder &   4.16 \\
        &               &                 Alcohol abuse with intoxication, unspecified &   3.13 \\
        &               &                                           Alcohol withdrawal &   3.09 \\ \cmidrule{2-4} 
        &           Brazilian & Motorcycle driver injured in collision with fixed or stat... &   3.45 \\ \cmidrule{2-4} 
        &      Other European &                  Obstructive sleep apnea (adult) (pediatric) &   1.46 \\
        &       &         Gastro-esophageal reflux disease without esophagitis &   1.31 \\
        &       &                                  Hyperlipidemia, unspecified &   1.30 \\
        &       &                             Essential (primary) hypertension &   1.30 \\
        &       &                      Personal history of nicotine dependence &   1.29 \\ \cmidrule{2-4} 
        &             Russian &    Unspecified hypertensive heart disease with heart failure &   9.06 \\
        &              &                                           Bifascicular block &   6.45 \\
        &              &                                 Nontoxic multinodular goiter &   4.73 \\
        &              &                                  Sinoatrial node dysfunction &   4.22 \\
        &              &                                         Unspecified glaucoma &   3.64 \\
\bottomrule
\end{tabular}

\caption{\textbf{Granular race groups exhibit significantly different patterns of ICD codes compared to their coarse groups.} We list ICD codes that are significantly enriched in a particular granular race group, as measured by the ratio of the prevalence of an ICD code among patients in a granular race to the prevalence of an ICD code in the remainder of the coarse subgroup. For example, the ICD code for ``Hypothyroidism" is 2.9 times more common among Indian patients compared to patients from other Asian subgroups. We apply a Bonferroni correction for multiple hypothesis testing and only report significantly enriched ICD codes. If a granular race group does not appear in the table, it is because no ICD code is significantly enriched in that patient subgroup. }
    \label{tab:enriched_icd_codes}

\end{table}

%% file: mlhc-camera-ready-template/table_enriched_eci_codes_fisher.tex
\begin{table}[]
    \centering
    \tiny

\begin{tabular}{llll}
\toprule
 Coarse Race &    Granular Race &                      ICD Code &  Ratio \\
\midrule
       Asian &      Indian &                Hypothyroidism &   3.14 \\ \cmidrule{2-4} 
        &          Chinese &         Tumor (w/ Metastasis) &   2.06 \\
        &           &        Tumor (w/o Metastasis) &   1.97 \\
        &           &                 Renal Failure &   1.96 \\
        &           &                  Coagulopathy &   1.94 \\
        &           &          Hypertension, Compl. &   1.93 \\ \cmidrule{2-4} 
        & SE Asian  &                   Weight Loss &   2.62 \\
        &  &         Chronic Pulm. Disease &   2.12 \\
        &  &          Hypertension, Compl. &   1.97 \\
        &  &        Tumor (w/o Metastasis) &   1.85 \\
        &  & Fluid \& Electrolyte Disorders &   1.66 \\ \midrule 
       Black & Black* &                    Drug Abuse &   2.54 \\
        &  &          Rheumatoid Arthritis &   2.20 \\
        &  &                 Alcohol Abuse &   2.11 \\
        &  &                       Obesity &   2.05 \\
        &  &         Chronic Pulm. Disease &   1.98 \\ \midrule 
Hisp./Latino &     Puerto Rican &                    Drug Abuse &   4.14 \\
 &      &         Chronic Pulm. Disease &   2.62 \\
 &      &                     Psychoses &   2.37 \\
 &      &                 Alcohol Abuse &   2.24 \\
 &      &                    Depression &   2.03 \\ \midrule 
       White &       Portuguese &                 Liver Disease &   2.53 \\
        &        &                  Coagulopathy &   2.00 \\
        &        &        Tumor (w/o Metastasis) &   1.84 \\
        &        &            Diabetes, Uncompl. &   1.72 \\ \cmidrule{2-4} 
        &           White* &                    Drug Abuse &   1.59 \\
        &            &                 Alcohol Abuse &   1.32 \\ \cmidrule{2-4} 
        &   Other European &        Tumor (w/o Metastasis) &   1.46 \\
        &    &         Tumor (w/ Metastasis) &   1.45 \\
        &    &                       Obesity &   1.31 \\
        &    &           Cardiac Arrhythmias &   1.21 \\
        &    &        Hypertension, Uncompl. &   1.21 \\ \cmidrule{2-4} 
        &          Russian &          Hypertension, Compl. &   2.82 \\
        &           &            Diabetes, Uncompl. &   2.82 \\
        &           &      Congestive Heart Failure &   2.72 \\
        &           &                 Renal Failure &   2.65 \\
        &           &        Hypertension, Uncompl. &   2.26 \\
\bottomrule
\end{tabular}
\caption{\textbf{Granular race groups are significantly enriched for certain ECI codes compared to the remaining patients in a coarse race group.}  We list ECI codes that are significantly enriched in a granular race group, as measured by the ratio of the prevalence of an ECI code among patients in a granular race to the prevalence of an ECI code in the remainder of the coarse subgroup. For example, the ECI code for ``Hypothyroidism" is 3.14 times more common among Indian patients compared to patients from other Asian subgroups. All p-values are computed with Bonferroni multiple hypothesis correction and are below .05, measured using a Fisher exact test for a difference in proportions. We exclude ECI codes which appear fewer than 10 times in a granular race group for privacy reasons. If a granular race does not appear here, it is because no ECI code is significantly enriched in that group.}
\label{tab:enriched_eci_codes}
\end{table}

%% file: mlhc-camera-ready-template/supp_pval_tables.tex
\begin{table}
\centering

\caption{\textbf{Granular variation in performance of the NEWS clinical risk score.} For each metric and coarse group, asterisks denote whether there is at least one granular group with significantly different predictive performance than the coarse group. All $p$-values are computed with Bonferroni multiple hypothesis correction. $\star$: $p < 0.05$, $\star\star$: $p < 0.01$, $\star\star\star$: $p < 0.001$, \texttt{-} not significant.}

\label{table:news}

\begin{tabular}{l|lcccc}
\toprule
        & \textbf{Metric} &              AUPRC &              AUROC &                FPR &           FNR \\
\textbf{Outcome} & \textbf{Coarse Race} &                    &                    &                    &               \\
\midrule
\multirow{4}{*}{\textbf{Hospitalization}} & \textbf{Asian} &  $\star\star\star$ &                  - &       $\star\star$ &             - \\
        & \textbf{Black} &  $\star\star\star$ &                  - &  $\star\star\star$ &             - \\
        & \textbf{Hispanic/Latino} &  $\star\star\star$ &                  - &  $\star\star\star$ &             - \\
        & \textbf{White} &  $\star\star\star$ &       $\star\star$ &  $\star\star\star$ &             - \\
\cline{1-6}
\multirow{4}{*}{\textbf{Critical}} & \textbf{Asian} &                  - &                  - &                  - &             - \\
        & \textbf{Black} &                  - &                  - &  $\star\star\star$ &             - \\
        & \textbf{Hispanic/Latino} &                  - &                  - &       $\star\star$ &             - \\
        & \textbf{White} &  $\star\star\star$ &  $\star\star\star$ &  $\star\star\star$ &  $\star\star$ \\
\cline{1-6}
\multirow{4}{*}{\textbf{Revisit}} & \textbf{Asian} &                  - &                  - &            $\star$ &             - \\
        & \textbf{Black} &  $\star\star\star$ &                  - &  $\star\star\star$ &             - \\
        & \textbf{Hispanic/Latino} &  $\star\star\star$ &                  - &  $\star\star\star$ &             - \\
        & \textbf{White} &  $\star\star\star$ &                  - &  $\star\star\star$ &             - \\
\bottomrule
\end{tabular}
\end{table}

\begin{table}
\centering

\caption{\textbf{Granular variation in performance of the CART clinical risk score.} For each metric and coarse group, asterisks denote whether there is at least one granular group with significantly different predictive performance than the coarse group. All $p$-values are computed with Bonferroni multiple hypothesis correction. $\star$: $p < 0.05$, $\star\star$: $p < 0.01$, $\star\star\star$: $p < 0.001$, \texttt{-} not significant.}

\label{table:cart}

\begin{tabular}{l|lcccc}
\toprule
        & \textbf{Metric} &              AUPRC &              AUROC &                FPR &                FNR \\
\textbf{Outcome} & \textbf{Coarse Race} &                    &                    &                    &                    \\
\midrule
\multirow{4}{*}{\textbf{Hospitalization}} & \textbf{Asian} &  $\star\star\star$ &                  - &  $\star\star\star$ &  $\star\star\star$ \\
        & \textbf{Black} &                  - &  $\star\star\star$ &  $\star\star\star$ &  $\star\star\star$ \\
        & \textbf{Hispanic/Latino} &  $\star\star\star$ &            $\star$ &  $\star\star\star$ &  $\star\star\star$ \\
        & \textbf{White} &  $\star\star\star$ &            $\star$ &  $\star\star\star$ &  $\star\star\star$ \\
\cline{1-6}
\multirow{4}{*}{\textbf{Critical}} & \textbf{Asian} &                  - &            $\star$ &  $\star\star\star$ &  $\star\star\star$ \\
        & \textbf{Black} &                  - &                  - &  $\star\star\star$ &                  - \\
        & \textbf{Hispanic/Latino} &                  - &                  - &  $\star\star\star$ &            $\star$ \\
        & \textbf{White} &                  - &                  - &  $\star\star\star$ &  $\star\star\star$ \\
\cline{1-6}
\multirow{4}{*}{\textbf{Revisit}} & \textbf{Asian} &                  - &  $\star\star\star$ &  $\star\star\star$ &  $\star\star\star$ \\
        & \textbf{Black} &  $\star\star\star$ &                  - &  $\star\star\star$ &                  - \\
        & \textbf{Hispanic/Latino} &  $\star\star\star$ &                  - &  $\star\star\star$ &                  - \\
        & \textbf{White} &  $\star\star\star$ &                  - &  $\star\star\star$ &  $\star\star\star$ \\
\bottomrule
\end{tabular}

\end{table}

\begin{table}
\centering

\caption{\textbf{Replicating granular variation results using XGBoost machine learning models.} This table replicates Table \ref{table:logreg_pval}, except uses results from a more complex machine learning model, instead of logistic regression. As mentioned in \S\ref{sec:methods_disparities}, the results are very similar to the results for the simpler logistic regression model discussed in the text.}

\label{table:xgb}
\begin{tabular}{l|lcccc}
\toprule
        & \textbf{Metric} &              AUPRC &              AUROC &                FPR &                FNR \\
\textbf{Outcome} & \textbf{Coarse Race} &                    &                    &                    &                    \\
\midrule
\multirow{4}{*}{\textbf{Hospitalization}} & \textbf{Asian} &  $\star\star\star$ &                  - &  $\star\star\star$ &  $\star\star\star$ \\
        & \textbf{Black} &                  - &  $\star\star\star$ &  $\star\star\star$ &            $\star$ \\
        & \textbf{Hispanic/Latino} &                  - &                  - &  $\star\star\star$ &            $\star$ \\
        & \textbf{White} &  $\star\star\star$ &            $\star$ &  $\star\star\star$ &  $\star\star\star$ \\
\cline{1-6}
\multirow{4}{*}{\textbf{Critical}} & \textbf{Asian} &                  - &                  - &  $\star\star\star$ &                  - \\
        & \textbf{Black} &                  - &                  - &  $\star\star\star$ &                  - \\
        & \textbf{Hispanic/Latino} &                  - &                  - &                  - &                  - \\
        & \textbf{White} &            $\star$ &                  - &  $\star\star\star$ &                  - \\
\cline{1-6}
\multirow{4}{*}{\textbf{Revisit}} & \textbf{Asian} &                  - &                  - &                  - &                  - \\
        & \textbf{Black} &  $\star\star\star$ &  $\star\star\star$ &  $\star\star\star$ &  $\star\star\star$ \\
        & \textbf{Hispanic/Latino} &       $\star\star$ &            $\star$ &  $\star\star\star$ &  $\star\star\star$ \\
        & \textbf{White} &  $\star\star\star$ &  $\star\star\star$ &  $\star\star\star$ &  $\star\star\star$ \\
\bottomrule
\end{tabular}
\end{table}

%% file: mlhc-camera-ready-template/table_feature_interactions.tex
\begin{table}[!ht]
\small
\centering
\caption{\textbf{Feature interactions with granular race for the critical outcome.} Significant $p$-values mean that the feature's coefficient varies within the corresponding coarse race group, i.e., there are granular differences in $p(y\mid x)$. The $p$-values shown are uncorrected, but we only display the $p$-value if it remains significant after a Bonferroni correction for the 120 hypothesis that are tested in this table (all displayed $p$-values are less than $0.05 / 120 = 0.00042$). For this analysis, we kept only the most predictive groups of features in order to restrict the number of tested hypotheses.}
\label{table:feature_interactions_crit}

\begin{tabular}{lllll}
\toprule
{} &    White & Black & Hispanic/Latino &    Asian \\
\midrule
age                &        - &                      - &                  - &        - \\
gender\_M           &        - &                      - &                  - &        - \\
n\_hosp\_365d        &        - &                      - &                  - &  3.2e-07 \\
n\_ed\_365d          &        - &                      - &                  - &        - \\
triage\_temperature &        - &                      - &                  - &        - \\
triage\_heartrate   &        - &                      - &                  - &  1.7e-05 \\
triage\_resprate    &  2.1e-06 &                      - &            4.5e-08 &        - \\
triage\_o2sat       &        - &                1.9e-07 &            0.00025 &        - \\
triage\_sbp         &        - &                      - &                  - &        - \\
triage\_dbp         &        - &                      - &                  - &        - \\
triage\_pain        &        - &                      - &                  - &        - \\
triage\_acuity      &  2.2e-08 &                1.9e-10 &                  - &  4.4e-06 \\
eci\_Arrhythmia     &        - &                      - &                  - &  2.7e-09 \\
eci\_Valvular       &        - &                      - &            0.00029 &        - \\
eci\_PHTN           &        - &                      - &                  - &        - \\
eci\_HTN1           &        - &                      - &            0.00025 &        - \\
eci\_HTN2           &        - &                      - &                  - &  3.4e-05 \\
eci\_NeuroOther     &        - &                0.00032 &                  - &  2.5e-09 \\
eci\_Hypothyroid    &        - &                      - &            0.00033 &        - \\
eci\_Lymphoma       &        - &                      - &                  - &        - \\
eci\_Coagulopathy   &        - &                      - &                  - &        - \\
eci\_Obesity        &        - &                      - &                  - &        - \\
eci\_WeightLoss     &        - &                      - &            0.00027 &        - \\
eci\_FluidsLytes    &        - &                0.00012 &                  - &  5.5e-06 \\
eci\_BloodLoss      &        - &                2.8e-05 &                  - &        - \\
eci\_Anemia         &        - &                7.8e-05 &                  - &  0.00036 \\
eci\_Alcohol        &        - &                      - &            7.5e-08 &  0.00019 \\
eci\_Drugs          &        - &                      - &            8.7e-06 &        - \\
eci\_Psychoses      &        - &                      - &                  - &        - \\
eci\_Depression     &        - &                5.6e-06 &                  - &        - \\
\bottomrule
\end{tabular}

\end{table}

\begin{table}[!ht]
\small
\centering
\caption{\textbf{Feature interactions with granular race for the hospitalization outcome.} Significant $p$-values mean that the feature's coefficient varies within the corresponding coarse race group, i.e., there are granular differences in $p(y\mid x)$. The $p$-values shown are uncorrected, but we only display the $p$-value if it remains significant after a Bonferroni correction for the 120 hypothesis that are tested in this table (all displayed $p$-values are less than $0.05 / 120 = 0.00042$). For this analysis, we kept only the most predictive groups of features in order to restrict the number of tested hypotheses.}
\label{table:feature_interactions_hosp}

\begin{tabular}{lllll}
\toprule
{} &    White &    Black & Hispanic/Latino &    Asian \\
\midrule
Age                           &        - &        - &               - &        - \\
Sex (True=Male)               &        - &        - &               - &        - \\
\# of HOSP visits within 365d  &    4e-09 &  4.2e-08 &         5.6e-25 &        - \\
\# of ED visits within 365d    &  9.4e-07 &  2.1e-06 &         2.1e-08 &        - \\
Temperature (C)               &        - &        - &               - &        - \\
Heartrate                     &        - &        - &               - &        - \\
Respiratory Rate              &        - &        - &               - &        - \\
Oxygen saturation (\%)         &        - &        - &               - &        - \\
Systolic Blood Pressure       &        - &        - &               - &        - \\
Diastolic Blood Pressure      &        - &        - &               - &        - \\
Triage Pain                   &        - &        - &               - &        - \\
Triage Severity Index         &        - &        - &               - &        - \\
Cardiac Arrhythmias           &  5.5e-08 &        - &               - &        - \\
Valvular Disease              &        - &        - &               - &        - \\
Pulmonary Circ. Disorders     &        - &        - &               - &        - \\
Hypertension, Compl.          &        - &  2.7e-05 &               - &  4.3e-05 \\
Hypertension, Uncompl.        &        - &  2.7e-05 &               - &        - \\
Other Neuro. Disorders        &        - &        - &               - &        - \\
Hypothyroidism                &        - &        - &               - &        - \\
Lymphoma                      &        - &        - &               - &        - \\
Coagulopathy                  &  4.7e-07 &        - &               - &        - \\
Obesity                       &        - &        - &               - &        - \\
Weight Loss                   &        - &        - &               - &        - \\
Fluid \& Electrolyte Disorders &        - &  2.9e-10 &         0.00011 &        - \\
Blood Loss Anemia             &        - &        - &               - &        - \\
Deficiency Anemia             &        - &        - &               - &        - \\
Alcohol Abuse                 &        - &        - &         3.7e-07 &        - \\
Drug Abuse                    &        - &        - &         1.9e-06 &        - \\
Psychoses                     &        - &  5.2e-05 &         7.4e-08 &        - \\
Depression                    &        - &        - &               - &        - \\
\bottomrule
\end{tabular}

\end{table}

%% file: mlhc-camera-ready-template/table_outcomes_scores_metrics.tex
% \begin{table}[!htb]

%     \caption{Overview of Outcomes, Metrics, and Scores}
%     \begin{minipage}{.3\linewidth}
%       \centering
      
%     \end{minipage}%
%     \begin{minipage}{.3\linewidth}
%       \centering
%         \caption{}
%     \end{minipage} 
%     \begin{minipage}{.3\linewidth}
%       \centering
%         \caption{}
%     \end{minipage} 
% \end{table}